%% file: main.tex
\documentclass[10pt,journal,compsoc]{IEEEtran}
\IEEEoverridecommandlockouts
\usepackage{cite}
\usepackage{amsmath,amssymb,amsfonts,bm}
\usepackage{algorithmic}
\usepackage{graphicx}
\usepackage{balance}
\usepackage{enumitem}
\usepackage{textcomp}
\usepackage{xcolor}
\usepackage{algorithm}
\usepackage{multirow}
\usepackage{url}
\usepackage{booktabs}
\usepackage{array}
\usepackage{caption}
\usepackage{colortbl}
\usepackage{balance}
\usepackage{float,color}
\usepackage{romannum}

\usepackage{subcaption}
\usepackage{stfloats}
\usepackage{tabularx}
\newcolumntype{Y}{>{\small\centering\arraybackslash}X}

\newtheorem{proof}{Proof}[section]
\def\BibTeX{{\rm B\kern-.05em{\sc i\kern-.025em b}\kern-.08em
    T\kern-.1667em\lower.7ex\hbox{E}\kern-.125emX}}
\begin{document}
\def\method{GDERec}

\newcommand{\R}[1]{{\color{black}{\sc} #1}}
\title{Learning Graph ODE for Continuous-Time \\Sequential Recommendation}

\author{Yifang Qin$^*$, Wei Ju$^*$,~\IEEEmembership{Member,~IEEE}, Hongjun Wu, Xiao Luo, and Ming Zhang% <-this % stops a space
\IEEEcompsocitemizethanks{
\IEEEcompsocthanksitem Yifang Qin, Wei Ju, Hongjun Wu, and Ming Zhang are with School of Computer Science, National Key Laboratory for Multimedia Information Processing,  Peking University, Beijing, China. (e-mail: qinyifang@pku.edu.cn, juwei@pku.edu.cn, dlmao3@163.com, mzhang\_cs@pku.edu.cn)
\IEEEcompsocthanksitem Xiao Luo is with Department of Computer Science, University of California, Los Angeles, USA. (e-mail: xiaoluo@cs.ucla.edu)
\IEEEcompsocthanksitem Yifang Qin and Wei Ju contribute equally to this work.
\IEEEcompsocthanksitem Corresponding authors: Xiao Luo and Ming Zhang}
% \IEEEcompsocthanksitem This paper is partially supported by the National Key Research and Development Program of China with Grant No. 2018AAA0101902 as well as the National Natural Science Foundation of China (NSFC Grant No. 62106008 and No. 62006004). 
}% <-this % stops a space
% \thanks{Manuscript received April 19, 2005; revised August 26, 2015.}}


\markboth{Journal of \LaTeX\ Class Files,~Vol.~14, No.~8, August~2015}%
{Shell \MakeLowercase{\textit{et al.}}: Bare Advanced Demo of IEEEtran.cls for IEEE Computer Society Journals}



\IEEEtitleabstractindextext{%
\begin{abstract}
\input{1_abstract}

\end{abstract}

}

\maketitle
\input{2_introduction}
\input{3_related_work}

\input{4_method}
\input{5_experiment}

\input{6_conclusion}

% is not selected <OR> if conference mode is selected - because compsoc
% conference papers position the abstract like regular (non-compsoc)
\IEEEdisplaynontitleabstractindextext

% \fi
\IEEEpeerreviewmaketitle

\ifCLASSOPTIONcompsoc
  \section*{Acknowledgments}
\else
  \section*{Acknowledgment}
\fi

The authors are grateful to the anonymous reviewers for critically reading this article and for giving important suggestions to improve this article. 

This paper is partially supported by National Key Research and Development Program of China with Grant No. 2023YFC3341203, the National Natural Science Foundation of China (NSFC Grant Numbers 62306014 and 62276002) as well as the China Postdoctoral Science Foundation with Grant No. 2023M730057.

\ifCLASSOPTIONcaptionsoff
  \newpage
\fi



%\bibliographystyle{IEEEtran}
% \balance
\bibliographystyle{IEEEtran}
\bibliography{rec}

\end{document}

%% file: 1_abstract.tex
Sequential recommendation aims at understanding user preference by capturing successive behavior correlations, which are usually represented as the item purchasing sequences based on their past interactions. Existing efforts generally predict the next item via modeling the sequential patterns. Despite effectiveness, there exist two natural deficiencies: (i) user preference is dynamic in nature, and the evolution of collaborative signals is often ignored; and (ii) the observed interactions are often irregularly-sampled, while existing methods model item transitions assuming uniform intervals. Thus, how to effectively model and predict the underlying dynamics for user preference becomes a critical research problem. To tackle the above challenges, in this paper, we focus on continuous-time sequential recommendation and propose a principled graph ordinary differential equation framework named \method{}. Technically, \method{} is characterized by an autoregressive graph ordinary differential equation consisting of two components, which are parameterized by two tailored graph neural networks (GNNs) respectively to capture user preference from the perspective of hybrid dynamical systems. On the one hand, we introduce a novel ordinary differential equation based GNN to \emph{implicitly} model the temporal evolution of the user-item interaction graph. On the other hand, an attention-based GNN is proposed to \emph{explicitly} incorporate collaborative attention to interaction signals when the interaction graph evolves over time. The two customized GNNs are trained alternately in an autoregressive manner to track the evolution of the underlying system from irregular observations, and thus learn effective representations of users and items beneficial to the sequential recommendation. Extensive experiments on five benchmark datasets demonstrate the superiority of our model over various state-of-the-art recommendation methods.
% The implementations are available at: https://anonymous.4open.science/r/GDERec/.

    % \method{} consists of two tailored graph neural networks (GNNs)

    % making it difficult to capture and predict the underlying dynamics.

%% file: 2_introduction.tex
\section{Introduction}

Recommender systems, as critical components to alleviate information overloading, have attracted significant attention for users to discover items of interest in various online applications such as e-commerce~\cite{wang2018billion,li2020hierarchical,zhang2022oa} and social media platforms~\cite{min2022divide,mousavi2022effective,song2022have}. The key of a successful recommender system lies in accurately predicting users’ interests toward items based on their historical interactions. Traditional recommendation methods such as matrix factorization~\cite{mnih2007probabilistic,gopalan2015scalable,koren2009matrix} usually hold the assumption of independence between different user behaviors. However, user preference is typically dynamically embedded in item transitions and sequence patterns, and successive behaviors can be highly correlated. One promising direction to effectively achieve this goal is the sequential recommendation (SR), which aims at explicitly modeling the correlations between successive user behaviors. The success of SR in the past few years have significantly enhanced user experience in both search efficiency and new product discovery.

To deeply characterize the sequential patterns of SR for dynamically modeling user preference, there are a large amount of methods have been proposed. Early works mainly focus on markov chains~\cite{rendle2010factorizing,he2016fusing} to model item transitions. To better explore the user preference of successive behaviors, researchers adopt recurrent neural networks (RNNs) and their variants~\cite{hochreiter1997long} to model the item purchasing sequences~\cite{hidasi2015session,quadrana2017personalizing,li2018learning}, due to their capability of capturing the long-term sequential dependencies. The recent success of Transformer~\cite{vaswani2017attention} also motivates the developments of a series of self-attention SR models~\cite{li2017neural,kang2018self,sun2019bert4rec,wu2020sse,fan2021continuous,fan2022sequential}. SASRec~\cite{kang2018self} is the pioneering work in leveraging Transformer for sequential recommendation, which employs self-attention mechanisms~\cite{vaswani2017attention} to adaptively assign weights to previous items. More recently, graph neural networks (GNNs)~\cite{kipf2017semi,luo2022clear,ju2023comprehensive,ju2023tgnn} have been widely adopted for enhancing existing SR methods~\cite{wu2019session,xu2019graph,yang2022multi}. The basic idea is to model the interaction data as graphs (e.g., the user-item interaction graph), and then learn effective representations of users and items for recommendation via propagating messages with user-item edges to capture high-order collaborative signals.

\begin{figure}[t]
    \centering
    \includegraphics[width=\linewidth]{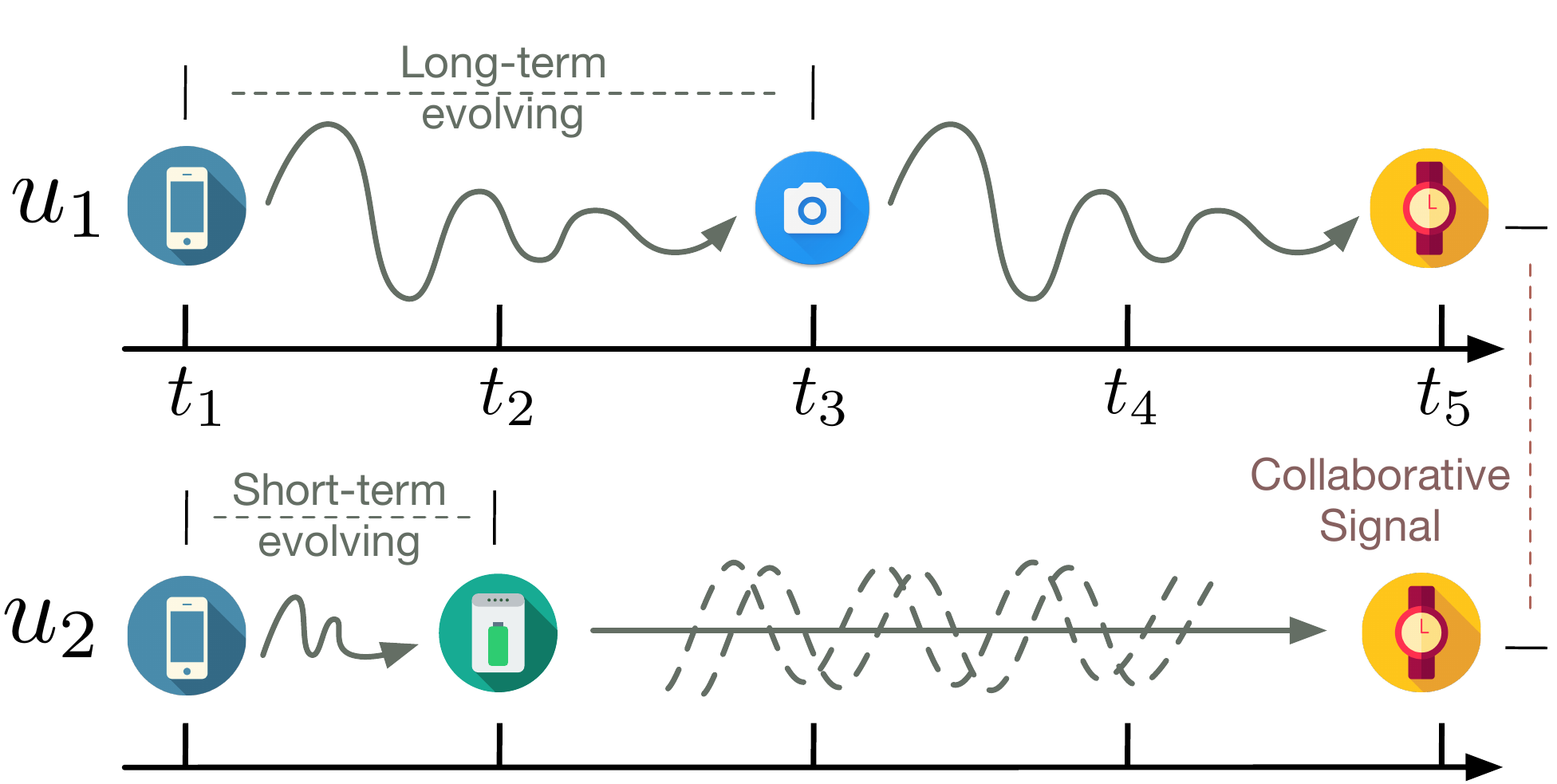}
    \caption{A toy example of observations sampled at different time intervals. Different time sampling intervals typically imply different user preferences.}
    \label{fig:illustration}
\end{figure}

Despite the encouraging performance achieved by these SR models, most existing approaches still suffer from two key limitations:
\textbf{(i) Inability to model the dynamic evolution of collaborative signals}. Actually, user preference is dynamic in nature. For example, a user may be interested in book items for a period of time and then search for new electronic games, and thus how to effectively model and understand the underlying dynamics becomes a critical problem. Moreover, the dynamic evolution of collaborative signals is also a crucial component in capturing user preference. Existing efforts generally predict the next item by merely modeling the item-item transitions inside sequences or capturing the interaction process in a static graph, ignoring the dynamic evolution of collaborative signals. \textbf{(ii) Fail to consider irregularly-sampled intervals of the observed interactions}. Most methods usually assume the observations are regularly sampled, which is impractical for many applications. As shown in Figure~\ref{fig:illustration}, user $u_1$ purchases ``phone$\rightarrow$camera$\rightarrow$watch'' at regular time intervals, we can speculate that he/she is a digital product enthusiast. However, user $u_2$ buys similar things at irregularly-sampled times, we may think that he/she buys a phone and power bank in a certain period of time because he/she just needs them, not because of his/her hobbies. The behavior of buying a watch much later, perhaps as a gift or for other reasons, suggests that irregularly-sampled observations could imply different user preferences. Therefore, how to explore the evolving process of successive user behaviors with irregular-sampled partial observations remains challenging. As such, we are looking for an approach that can model the dynamic evolution of collaborative signals and meanwhile overcome the irregularly-sampled observations.

Having realized the above challenges with existing SR methods, we focus on continuous-time sequential recommendation. Towards this end, this work proposes a principled graph ordinary differential equation framework named \method{}. The key idea of \method{} is to simultaneously characterize the evolutionary dynamics and adapt to irregularly-sampled observations. To achieve this goal effectively, \method{} is modeled by an autoregressive graph ODE composed of two components, which are parameterized by two tailored GNNs respectively to capture user preference from a hybrid dynamical systems view. On the one hand, we develop an ODE-based GNN to \emph{implicitly} capture the temporal evolution of the user-item interaction graph. On the other hand, we design an attention-based GNN to \emph{explicitly} incorporate collaborative attention to interaction signals when the interaction graph evolves over time. Further, the whole process can be optimized by alternately training two customized GNNs in an autoregressive manner to track the evolution of the underlying system from irregular observations. Our experiments show that it can largely improve the existing state-of-the-art approaches on five benchmark datasets. In a nutshell, we summarize the main contributions of this work are as follows:

\begin{itemize}[leftmargin=*]
\item \textbf{General Aspects:} Different from existing works in recommendation in discrete states, we explore continuous-time sequential recommendation, which remains under-explored and is able to generalize to any unseen times-tamps for future interactions. 
\item \textbf{Novel Methodologies:} We propose a novel framework to alternately train two parts of our autoregressive graph ODE in a principled way, capable of modeling the evolution of collaborative signals and overcoming irregularly-sampled observations.
\item \textbf{Multifaceted Experiments:} We conduct comprehensive experiments on five benchmark datasets to evaluate the superiority as well as the functionality of the proposed approach against competitive baselines.
\end{itemize}

% , which is still insufficient to yield satisfactory results

% Thus, user preference is inherently an evolving process, and 

% The rest of this paper is organized as follows. 
% Section~\ref{sec::related} reviews related work and Section~\ref{sec::definition} gives the problem definition as well as introduces some preliminaries. In Section~\ref{sec::model}, we describe the details of our proposed framework. Experiments on several benchmark datasets are presented in Section~\ref{sec::experiment} with discussions. Finally, Section~\ref{sec::conclusion} concludes the paper and visions for the future work.

%% file: 3_related_work.tex
\section{Related Work}
\label{sec::related}

In this section, we briefly review the related works in three aspects, namely sequential recommendation, graph neural networks, and ordinary differential equation.

\subsection{Sequential Recommendation}

Sequential recommendation (SR) aims to predict the next item based on the user’s historical interactions as a sequence by sorting interactions chronologically. The effectiveness of this method makes it possible to provide more timely and accurate recommendations. Existing approaches for SR can be mainly divided into markov chains (MC)~\cite{rendle2010factorizing,he2016fusing}, RNN-based~\cite{hidasi2015session,quadrana2017personalizing,li2018learning}, and Transformer-based~\cite{li2017neural,kang2018self,sun2019bert4rec,wu2020sse,fan2021continuous,fan2022sequential}. Most traditional methods are based on MC and the main purpose is to model item-to-item transaction patterns. For example, FPMC~\cite{rendle2010factorizing} regards user behaviors as markov chains, and estimates the user preference by learning a transition graph over items. With the recent advances of deep learning, many deep SR models leverage RNNs to capture long-term sequential dependencies. For instance, GRU4Rec~\cite{hidasi2015session} leverages RNNs to incorporate more abundant history information for the session-based recommendation. The success of Transformer~\cite{vaswani2017attention} inspires the adoption of the attention mechanism into SR. BERT4Rec~\cite{sun2019bert4rec} further applies deep bidirectional self-attention to model user behavior sequences. Different from these methods, our \method{} steps further and focuses on under-explored continuous-time SR, while existing methods fail to generalize to any unseen future times-tamps.

\subsection{Graph Neural Networks}

With the powerful capability of processing non-euclidean structured data, graph neural networks (GNNs)~\cite{kipf2017semi,qin2023disenpoi,luo2022dualgraph,ju2023glcc,ju2023unsupervised} have achieved wide attention due to their remarkable performance. The underlying idea is to update node representations using a combination of the current node’s representation and that of its neighbors following message passing schemas~\cite{gilmer2017neural}. Recently, GNNs have shown great promise for enhancing existing SR methods~\cite{wu2019session,song2019session,xu2019graph,yang2022multi,wang2022disenctr,ju2022kernel,qin2023diffusion}. SR-GNN~\cite{wu2019session} makes the first attempt to incorporate GNNs into the SR, and models session sequences as the graph to capture complex transitions of items. DGRec~\cite{song2019session} leverages graph attention neural network~\cite{velivckovic2018graph} to dynamically estimate the social influences based on users’ current interests. However, most of the GNN-based recommendation methods only focus on static graph scenarios, and have shown the inability to track the evolution of collaborative signals and overcome irregularly-sampled observed interactions, while our \method{} innovatively addresses these limitations.

\begin{figure*}[t!]
    \centering
    \includegraphics[width=\textwidth]{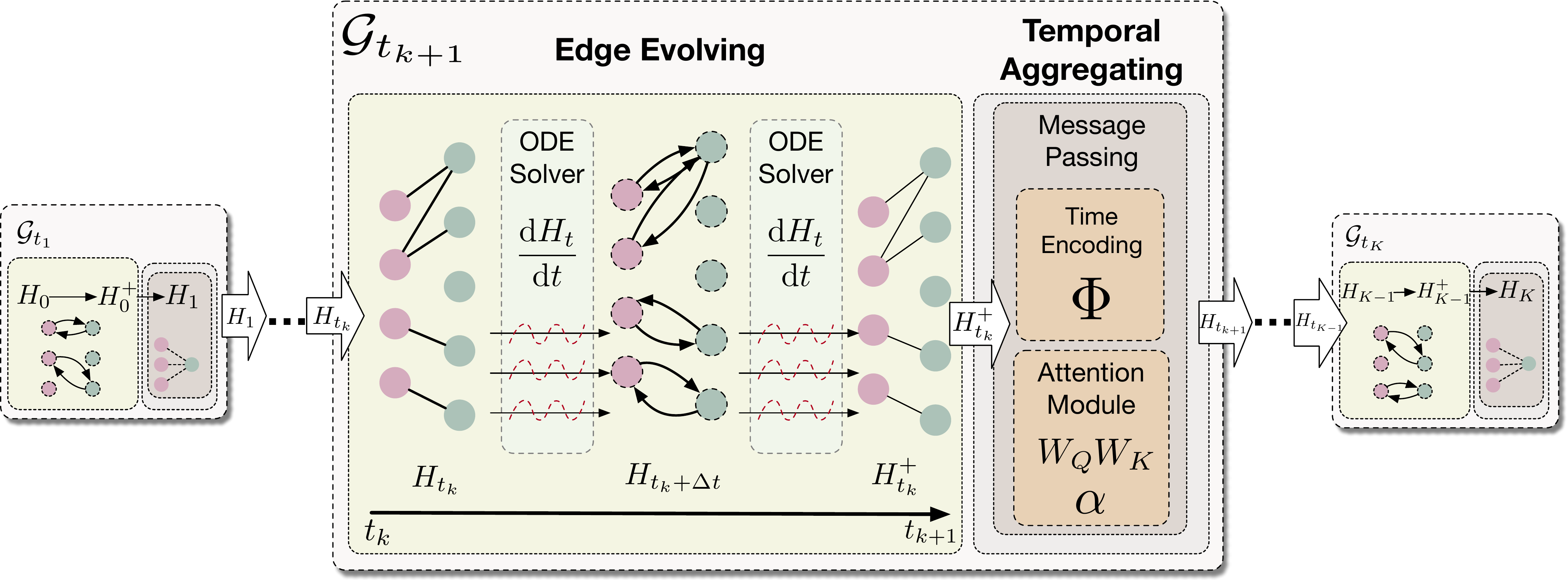}
    \caption{Illustration of the proposed framework \method{}. We propose an autoregressive framework that propagates on a hybrid dynamic interaction system. A basic unit of our framework is composed of two modules. The fed node representations from previous layers are first processed via an ODE-based edge evolving module to generate $H_{t_k}^+$. Then a temporal attention module aggregates neighborhood information to generate $H_{t_{k+1}}$ to be fed into the next layer.
    % We first generate two positive views by two pooling modules. Then, the two augmented views are fed into the online network while the original graph is fed into the target network. Our contrastive learning framework contains similarity learning and consistency learning, where the pooling modules are adversarially trained with respect to the encoder.
    }
    \label{fig:framework}
\end{figure*}

\subsection{Ordinary Differential Equation}

Neural ODE~\cite{chen2018neural} has been proposed as a new paradigm for generalizing discrete deep neural networks to continuous-time scenarios. It forms a family of models that approximate the ResNet~\cite{he2016deep} architecture by using continuous-time ODEs. Due to the superior performance and flexible capability, neural ODEs have been widely adopted in various research fields, such as traffic flow forecasting~\cite{fang2021spatial,choi2022graph,ji2022stden,rao2022fogs}, time series forecasting~\cite{chen2011time,de2019gru,jin2022multivariate}, and continuous dynamical system~\cite{huang2020learning,huang2021coupled,luo2023hope}. Recently, there are some advanced methods connecting GNNs and neural ODEs~\cite{zhuang2019ordinary,xhonneux2020continuous}. GODE~\cite{zhuang2019ordinary} generalizes the concept of continuous-depth models to graphs, and parameterizes the derivative of hidden node states with GNNs. Compared with existing ODE methods, our work goes further and extends this idea to investigate an under-explored yet important continuous-time sequential recommendation.

% CGNN~\cite{xhonneux2020continuous} characterizes the derivatives of node representations w.r.t. time to derive continuous dynamics of node representations.

%% file: 4_method.tex
\section{Preliminary}
\label{sec::definition}

As the fundamental recommendation problem, sequential recommendation system aims to predict the preference of users based on the observed historical user-item interaction sequences. Graph neural network (GNN) based recommendation approaches have been proposed to represent the interaction between users and items as a bipartite graph. Specifically, let $\mathcal U$ and $\mathcal I$ be the sets of users and items respectively, the temporal interaction bipartite graph is formulated as $\mathcal{G}=\{\mathcal{V},\mathcal{E}\}$, where $\mathcal{V}=\{\mathcal{U}\cup\mathcal{I}\}$ denotes the vertices set and $\mathcal{E}=\{(u,i,t)|u \text{ interacted with } i \text{ at time } t\}$ is the temporal edge set denoting the observed interactions.

Since the observed interactions $\mathcal{E}$ have taken place in the temporal order, given a set of pivot timestamps $\mathcal{T}=\{t_k\}_{k=1}^K,K\in\mathbb{N}^+$, we can consequently define the \emph{hybrid time domain} $\mathcal{I}=\bigcup_{k=1}^K([t_k,t_{k+1}],k)$ by dividing the continuous time domain into different slices separated by the pivot timestamps. The interaction stream on the graph can be further viewed as a \emph{hybrid dynamic system} $\{\mathcal{G}_{t_k}\}_{k=1}^K$, where each 
\begin{equation}
\label{eq:graph}
    \mathcal{E}_{t_k}=\{e=(u,i,t)|e\in\mathcal{E}\land t\in[t_k,t_{k+1}]\}\}.
\end{equation}

The construction of $\mathcal{G}_{t_k}$ depicts the hybrid dynamic process of a given interaction system by representing the continuous evolution process \emph{between} pivot timestamps and the discrete instant influence of interactions \emph{at} each pivot $t_k$. Since we have constructed the system, the object of GNN-based sequential recommendation is to predict the interactions during next period $[t_K,t_{K+1}]$, given the interaction graph $\mathcal{G}$. Additionally, to fully leverage the hybrid dynamic of the system, the method is expected to be capable of handling both the continuous evolution within time slices and the sudden changes at discrete time pivots.

\section{Methodology}
\label{sec::model}

\subsection{Overview}
\label{sec::overview}

In this paper, we introduce our approach for continuous-time sequential recommendation by incorporating the evolutionary dynamics of underlying systems. Existing methods merely explore the sequential patterns and model item transitions assuming uniform intervals. However, they are not able to fully capture the evolution of collaborative signals and fail to adapt to irregularly-sampled observations, which leads to insufficient expressiveness.

We address the above limitations by proposing a novel graph ordinary differential equation framework called \method{}. Specifically, \method{} builds a principled autoregressive graph ODE consisting of two parameterized GNNs. On the one hand, a tailored ODE-based GNN is developed to implicitly track the temporal evolution of collaborative signals on the user-item interaction graph. On the other hand, we leverage the attention mechanisms to equip GNN to explicitly capture the evolving interactions when the interaction graph evolves over time. The two components are trained alternately to learn effective representations of users and items beneficial to the recommender system. An illustration of the framework is presented in Figure~\ref{fig:framework}. Next, we will introduce the edge evolving module and the temporal aggregating module. Finally, the training algorithm to optimize the model is explained.

\subsection{Edge Evolving Module}
GNNs propagate and aggregate messages on given graph adjacency structures. Typically, graph convolution layers take the form of the following message passing scheme:
\begin{equation}
    H_{l+1}=GCN(H_l)=AH_l,
\end{equation}
where $H_l\in\mathbb{R}^{|\mathcal{V}|\times d}$ denotes the hidden representation of nodes on the $l$-th layer and $H_0$ is the initial node embeddings. $A$ represents the normalized graph adjacency matrix. The message function of GCNs weights the influence of neighborhood by the normalized node degrees, which ignores the natural affinity between connected nodes. However, the user would receive different influences from the interacted items. Towards this end, we follow the advice of previous work \cite{wang2019neural} and rewrite the propagation process:
\begin{equation}
\label{eq:prop}
    H_{l+1}=AH_l+AH_0\odot H_0,
\end{equation}
\R{where the first item in the message function represents the standard message passed via graph convolution, while the second item takes the form of the element-wise product of the source and the destination nodes' representations to model the affinity between neighboring nodes.}

Specifically, in recommendation scenarios, user preference and item representation would evolve with the continuous-time flow. In other words, it requires the extent of the discrete propagation to continuous form to further simulate the temporal evolution process on a dynamic graph. Intuitively, we can expand Eq.~\ref{eq:prop} as:
\begin{equation}
\label{eq:sum}
    \begin{aligned}
    H_l &=A^lH_0+(\sum_{i=1}^lA^i)H_0\odot H_0 \\
     &=A^lH_0+(A-I)^{-1}(A^{l+1}-A)H_0\odot H_0.
    \end{aligned}
\end{equation}

From Eq. \ref{eq:sum} we can obtain the closed-form solution of neighborhood propagation with arbitrary number of layers. To get more fine-grained representations of the interaction graph, we hope to extend Eq. \ref{eq:sum} to continuous form. Specifically, we replace the discrete $l$ with a continuous variable $t$ and Eq. \ref{eq:sum} can thus be viewed as a Riemann sum. We have the integral formulation:
\begin{equation}
\label{eq:reimann}
    H_t=A^tH_0+\int_{0}^{t+1}A^\tau H_0\odot  H_0\text{d}\tau-H_0\odot H_0.
\end{equation}

However, the item $A^t$ is intractable to compute for $t\in\mathbb{R}$. Therefore we aim to reform the calculation of $H_t$ as an ordinary differential equation and state the following propositions to elicit the numerical solution of $H_t$.
%into an ODE structure by considering its derivative:

\smallskip\textbf{Proposition 1} \emph{The first-order derivative of $H_t$ in Eq. \ref{eq:reimann} can be formulated as the following ODE:}
\begin{equation}
    \frac{\text{d}H_t}{\text{d}t}=\ln AH_t+AH_0\odot H_0,
\end{equation}
\emph{where the initial $H_0$ is the output from downstream networks.}
\begin{proof}
    We first directly calculate the derivative of $H_t$ as:
\begin{equation}
\label{eq:1st-order}
    \frac{\text{d}H_t}{\text{d}t}=\ln AA^tH_0+A^{t+1}H_0\odot H_0.
\end{equation}
To get rid of the remaining items with $A^t$, We further calculate the second-order derivative of $H_t$:
\begin{equation}
\label{eq:2nd-order}
\begin{aligned}
    \frac{\text{d}^2H_t}{\text{d}t^2}&=\ln^2 AA^tH_0+\ln AA^{t+1}H_0\odot H_0 \\
    &=\ln A\frac{\text{d}H_t}{\text{d}t}.
\end{aligned}
\end{equation}
By integration on both sides of Eq. \ref{eq:2nd-order}, we have:
\begin{equation}
\label{eq:integral}
    \frac{\text{d}H_t}{\text{d}t}=\ln AH_t+const.
\end{equation}
To solve the value of $const$, we let $t=0$ and combine Eq. \ref{eq:1st-order} and \ref{eq:integral}:
\begin{equation}
    \frac{\text{d}H_t}{\text{d}t}\Big|_{t=0}=\ln AH_0+AH_0\odot H_0=\ln AH_0+const.
\end{equation}
We get that:
\begin{equation}
    const=AH_0\odot H_0,
\end{equation}
and the derivative form of the ODE process can be rewritten as:
\begin{equation}
\label{eq:ode}
    \frac{\text{d}H_t}{\text{d}t}=\ln AH_t+AH_0\odot H_0. \\
\end{equation}
\end{proof}

So far we have obtained the first order derivative of $H_t$ with respective to time variable $t$ that is only determined by the value of $H_t$, the initial representation matrix $H_0$, and the normalized adjacency $A$. The formulation of Eq. \ref{eq:ode} can be further fed into neural ODE frameworks \cite{chen2018neural} to model the evolving process of $H_t$.

To calculate $\ln A$ in practice, we approximate it by making a first-order Taylor approximation to $\ln A$.
\begin{equation}
\label{eq:taylor}
    \frac{\text{d}H_t}{\text{d}t}=(A-I)H_t+AH_0\odot H_0.
    % \left.\frac{\text{d}H_t}{\text{d}t}\right|_{t=0}=(A-I)H_t+AH_0\odot H_0,
\end{equation}

Specifically, the ODE we propose has an analytical solution.

\smallskip\textbf{Proposition 2} \emph{The analytical solution of Eq. \ref{eq:taylor} is given by:}
\begin{equation}
    H_t=(A-I)^{-1}((e^{(A-I)t}-I)AH_0\odot H_0)+e^{(A-I)t}H_0
\end{equation}
\begin{proof}
    To solve the ODE defined by Eq. \ref{eq:taylor}, we first multiply both sides of the equation by an exponential factor $\exp(-(A-I)t)$ and rearrange the items:
    \begin{equation}
        e^{-(A-I)t}\frac{\text{d}H_t}{\text{d}t}-e^{-(A-I)t}(A-I)H_t=e^{-(A-I)t}AH_0\odot H_0.
    \end{equation}
    By integrating from 0 to a specified $\tau$ on both side, we can thus obtain $H_{\tau}$ by the integral results:
    \begin{equation}
    \label{eq:int_result}
        e^{-(A-I)t}H_t\Big|_{t=0}^\tau+S-S=-(A-I)^{-1}e^{-(A-I)t}AH_0\odot H_0\Big|_{t=0}^\tau,
    \end{equation}
    where the item S is given by:
    \begin{equation}
        S = \int_{0}^\tau(e^{-(A-I)t}(A-I)H_t)dt.
    \end{equation}
    From Eq. \ref{eq:int_result} we obtain $H_t$ for any $t>0$:
    \begin{equation}
    \begin{aligned}
        &e^{-(A-I)t}H_t-H_0=\\
        &-(A-I)^{-1}(e^{-(A-I)t}AH_0\odot H_0-AH_0\odot H_0),
    \end{aligned}
    \end{equation}
    where the analytical solution of $H_t$ is given by:
    \begin{equation}
        H_t=(A-I)^{-1}((e^{(A-I)t}-I)AH_0\odot H_0)+e^{(A-I)t}H_0
    \end{equation}
\end{proof}
Since we have modeled the discrete dynamics in Eq. \ref{eq:taylor} of evolving interacting system in an ODE formulation, the value of $H_t$ given a continuous time $t$ can then be solved by a designated ODE solver such as \emph{Runge–Kutta} method \cite{runge1895numerische}:

\begin{equation}
    H_t=\text{ODESolver}(\frac{\text{d}H_t}{\text{d}t},H_0,t)
    \label{eq:ode_solver}
\end{equation}

\R{For fixed step size $\epsilon$, a typical ODE solver would iteratively update the value of $H_t$. For instance, to solve an initial value problem of ODEs that formulated as:
\begin{equation}
    H_T=H_0+\int_0^T f(H_t)dt,
\end{equation}
where $f(H_t)$ represents $\frac{\text{d}H_t}{\text{d}t}$ in our case. The Runge-Kutta-4 (RK-4) solver updates the value with:
\begin{equation}
\begin{aligned}
    k_i&=\begin{cases}
      f(H_t) & i=1 \\
      f(H_t+\frac{\epsilon}{2}k_1) & i=2 \\
      f(H_t+\frac{\epsilon}{2}k_2) & i=3 \\
      f(H_t+\epsilon k_3) & i=4 
    \end{cases} \\
    H_{t+\epsilon}&=H_t+\frac{\epsilon}{6}(k_1+2k_2+2k_3+k_4) \\
\end{aligned}
\end{equation}
}   
\subsection{Temporal Aggregating Module}
Recalling that the constructed hybrid dynamic system in Eq. \ref{eq:graph} is composed of the continuous intervals between pivot times and the discrete pivot timestamps. While the edge evolving module models the continuously evolving process between any $t_k$ and $t_{k+1}$, we design a temporal aggregating module to explicitly aggregate neighboring information on a given interaction graph. For instance, given the temporal edges $\mathcal{E}_{t<t_{k+1}}$, which represents the interactions before $t_{k+1}$, and the node representations $H_{t_k}^+$ calculated in Eq. \ref{eq:ode_solver}, our goal is to obtain a propagation function $\textbf{F}$:
\begin{equation}
    H_{t_{k+1}}=\textbf{F}(\mathcal{E}_{t<t_{k+1}},H_{t_k}^+,\Theta_k),
\end{equation}
where $H_{t_{k+1}}$ is the output representation of the current layer, $\Theta$ denotes trainable network parameters.

\subsubsection{Trainable Time Encoding}
To explicitly encode temporal information presented in the temporal edges, we propose to introduce a time mapping $\Phi:T\rightarrow\mathbb{R}^{d_T}$ that maps a given timestamp into the latent space. Previous researches \cite{xu2019self} proposed several \emph{translation-invariant} time encoding functions, i.e., $\Phi$ that satisfy:
\begin{equation}
    \left<\Phi(t_1),\Phi(t_2)\right>=\left<\Phi(t_1+c),\Phi(t_2+c)\right>,\forall c\in\mathbb{R},
\end{equation}
where $\left< \cdot, \cdot\right>$ denotes the inner product of two given vectors. Here we implicate Bochner's theorem \cite{loomis2013introduction} as our time encoding function:
\begin{equation}
    \Phi(t):=\sqrt{\frac{1}{d_T}}[\cos(\omega_1t),\sin(\omega_1t),...,\cos(\omega_{d_T}t),\sin(\omega_{d_T}t)]^\top,
\label{eq:temporal_encode}
\end{equation}
where $\mathbf{\omega}=[\omega_1,\omega_2,...,\omega_{d_T}]\in\mathbb{R}^{d_T}$ is the trainable time embedding to reflect flexible temporal characteristics.

\subsubsection{Temporal Attention Network}
Graph attention networks~\cite{velivckovic2018graph} suggest leveraging the target attention mechanism in the message function. To incorporate temporal factors into the attention mechanism, we propose to build an attention network that jointly captures chronological and contextual information. Particularly, given the hidden representation $H$ of the previous layer and temporal edges $\mathcal{E}$, \R{the message passed from arbitrary node $j$ to node $i$ at $k$-th layer is constructed as:}
\begin{equation}
    h_i=\sum_{j\in\mathcal{N}_i}m_{i\leftarrow j}(t)=\frac{1}{\sqrt{|\mathcal{N}_i||\mathcal{N}_j|}}\pi_{t}(i,j)h_j^+,
\end{equation}
\R{where $m_{i\leftarrow j}$ represents the passed message}. For target node $i$, $j\in\mathcal{N}_i$ denotes the neighboring node of $i$ in $\mathcal{E}_{t<t_{k+1}}$. We follow GCNs \cite{kipf2017semi} and apply the Laplacian normalization to each node to avoid the over-squashing issue on the large user-item graph. The attention weight $\pi_t(i,j)$ is calculated via the soft-attention mechanism to model the characterized contribution from the neighborhood to each node:
\begin{equation}
    \pi_t(i,j)=\sigma(\alpha^\top\cdot \text{CONCAT}(W_Qh_i^{(l)};\Phi(t);W_Kh_j^{(l)})),
\end{equation}
where $\alpha\in\mathbb{R}^{2d+d_T},W_Q,W_K\in\mathbb R^{d\times d}$ are all trainable attention parameters, $\sigma(\cdot)$ denotes sigmoid function:
\begin{equation}
    \sigma(x)=\frac{1}{1+e^{-x}}.
\end{equation}

\subsection{Autoregressive Propagation Module}
Since we have defined the formulation of the edge evolving module and the temporal aggregating module, given an interaction graph $\mathcal{G}$, we can build the corresponding hybrid dynamic interacting system $\{\mathcal{G}_{t_k}\}_{k=1}^K$ and the initial node embeddings $H_{t_0}$. We design an autoregressive framework that propagates messages on the hybrid dynamic time domain. To formulate, we have:
\begin{equation}
\begin{cases}
\label{eq:regressive}
    H^+_{t_k}&=\text{ODESolver}(\frac{\text{d}H_t}{\text{d}t},H_{t_k},t_{k+1}-t_k) \\
    H_{t_{k+1}}&=\textbf{F}(\mathcal{E}_{t<t_{k+1}},H_{t_k}^+,\Theta_{k}),
\end{cases}
\end{equation}
where $\{\Theta_k\}_{k=1}^K$ are list of model parameters. 

So far we have defined the whole propagation process on a given hybrid dynamic interacting system, where the graph ODE-based module steers the evolving dynamics of the interacting process and calculates $H_{t_k}^+$ for each layer, and then the message-passing-based module explicitly models the temporal factors via a tailored temporal attention mechanism to output $H_{t_{k+1}}$ for next period of time. \method{} obtains layers of hidden representations by applying these two modules autoregressively as formulated in Eq. \ref{eq:regressive}.

\subsection{Model Prediction and Optimization}
After iteratively propagating along the hybrid dynamic interacting system $\{\mathcal{G}_{t_k}\}_{k=1}^K$, we can obtain the hidden representation of nodes on each layer, namely $H_{t_0},H_{t_1},...,H_{t_K}$, where for each layer we have $H_{t_k}=[h_{1,t_k},h_{2,t_k},...,h_{|\mathcal{V}|,t_k}]$. For a given user-item pair $(u,i)$, we obtain the corresponding node representation via:
\begin{equation}
    h_u=\frac{1}{K}\sum_{k=0}^Kh_{u,t_k},\ h_i=\frac{1}{K}\sum_{k=0}^Kh_{i,t_k}.
\end{equation}
% In order to jointly consider the node representation at every pivot time, we exploit both temporal and interacting features for each node on the interaction graph. 
To model the preference for the designated user $u$ to target item $i$, we calculate the rating score by the inner product of representations:
\begin{equation}
    \hat y_{ui}=h_u^\top h_i.
\end{equation}
To optimize the model parameters, we adapt Bayesian Personalized Ranking (BPR)~\cite{rendle2012bpr} loss as the target function, formulated as:
\begin{equation}
    \mathcal{L}_{BPR}=\sum_{(u,i,j)\in\mathcal{O}}-\ln \sigma(\hat y_{ui}-\hat y_{uj})+\lambda \lVert\Theta\rVert_2^2,
\end{equation}
where $\Theta$ denotes the involved model parameters and node embeddings, $\mathcal{O}=\{(u,i,j)|(u,i)\in\mathcal{E},(u,j)\in\mathcal{O}^-\}$. For each iteration, we randomly sample the negative interaction set $\mathcal{O}^-$ from $\mathcal{U}\times\mathcal{I}-\mathcal{E}$.

\smallskip\textbf{Complexity Analysis.}
Following the autoregressive propagation schema defined in previous sections, the computational consumption is mainly composed of two parts: (i) the ODE solver that estimates the solution of the proposed graph ODEs; (ii) the graph propagation process that aggregates node features on graphs.

For (i), there are several kinds of numerical methods to solve a given ordinary differential equation, for example, fixed-step methods like \emph{Runge-Kutta} method \cite{runge1895numerische} and adaptive-step methods like \emph{Dormand-Prince-Shampine} method \cite{dormand1980family}. In this work, we choose the \emph{Runge-Kutta-s} algorithm to solve the ODE with a fixed step of $s$-th order. Thus for a fixed step size $\epsilon$, the total time complexity of the solving process can be calculated via:
\begin{equation}
    O(\sum_{k=0}^K\frac{s}{\epsilon}d|\mathcal{E}|_{t_k})=O(\frac{s}{\epsilon}d|\mathcal{E}|).
\end{equation}

For (ii), the complexity of the edge propagation is only determined by the number of edges and the size of node representation:
\begin{equation}
    O((2d+d_t)|\mathcal{E}|).
\end{equation}

To sum up, we have the overall complexity of \method{}:
\begin{equation}
    O(((2+\frac{s}{\epsilon})d+d_t)|\mathcal{E}|)
\end{equation}

\subsection{\R{Comparison with Existing Models}}

\R{As a graph-based ODE model, \method{} is related to several existing methods that includes:}
\begin{itemize}[leftmargin=*]
    \item \R{LightGCN \cite{he2020lightgcn}: LightGCN is a classical graph convolution-based approach that conducts normalized graph convolution on the user-item interaction graph. Nonetheless, it overlooks the temporal disparities among interactions and the evolution of user interests. In contrast, our \method{} extends the utility of interaction graphs by introducing the concept of a hybrid dynamic system and a graph ODE network, which enable us to model the continuous evolution of user interests. }
    \item \R{GNG-ODE \cite{guo2022evolutionary}: GNG-ODE is a session graph-based recommendation method that integrates neural ODEs with session graphs. Specifically, GNG-ODE captures the evolving dynamics of users by proposing the t-Alignment technique, which enables the session graph to expand over time, which is similar to our proposed hybrid dynamic system. However, GNG-ODE only performs graph encoders on session graphs,  thereby neglecting the evolving collaborative signals between users and items. In contrast, our \method{} enhances the interaction system by incorporating time-evolving interaction edges. This hybrid dynamics facilitates the model's reception of more finely-grained global collaborative signals.}
    % which ignores the evolving collaborative signals between users and items. In addition, \method{} equips the graph encoders with the time-evolving interaction edges that allows the model to receive more finely-grained global collaborative signals.}
\end{itemize}

\R{In summary, our \method{} stands out as the first endeavor to integrate interaction dynamics into continuous-time recommendations, leveraging the novel hybrid dynamic systems and an ODE-based autoregressive propagation approach. The subsequent sections will provide empirical evidence of the effectiveness of our \method{}}

%% file: 5_experiment.tex
\section{Experiment}

We conduct comprehensive experiments on several benchmark datasets to evaluate the effectiveness of the proposed \method{} and answer the following research questions.

\textbf{RQ1: }
How does the proposed \method{} perform on real-world datasets compared with the current state-of-art methods? Is the idea of modeling hybrid dynamics effective for recommendations?

\textbf{RQ2: }
How does the idea of the hybrid dynamic interaction system enhance the recommendations? How do the hyper-parameters influence \method{}'s performance?

\textbf{RQ3: }
Can \method{} capture dynamic features of the interactions effectively as time flows? Is there a way to visualize the learned dynamics system in the hidden space?

\subsection{Overall Comparison (RQ1)}
\subsubsection{Datasets and Experimental Setup}

\begin{table}
\centering
\caption{Descriptive statistics of the used datasets.}
\label{tab:statics}
\setlength{\tabcolsep}{3pt}
\begin{tabular}{ccccc} 
\toprule %\hline
\textbf{Dataset} & \#User & \#Item & \#Interactions & Density \\
\midrule
Cloth & 39,387 & 23,033 & 278,677 & 0.31\% \\
\midrule
Baby & 19,445 & 7,050 & 160,792 & 0.11\% \\
\midrule
Music & 5,541 & 3,568 & 64,706 & 0.32\% \\
\midrule
ML-1M & 6,040 & 3,706 & 1,000,209 & 4.46\% \\
\midrule
ML-100K & 943 & 1,682 & 100,000 & 6.30\% \\
\bottomrule %\hline
\end{tabular}
\end{table}

We evaluate \method{} and baseline methods on five real-world datasets, including three subsets of the Amazon review dataset\footnote{{\url{https://jmcauley.ucsd.edu/data/amazon/}}}, namely Electronics, Cloth, and Music, and two subsets of the Movie-Lens dataset\footnote{{\url{https://grouplens.org/datasets/movielens/}}}, namely ML-1M and ML-100K. The detailed statistics of the used datasets are presented in table \ref{tab:statics}. The observed interactions in each dataset are chronologically sorted by the timestamp and then split into train/valid/test sets by an 80\%/10\%/10\% ratio, which is a common practice.

For \method{}, we construct the hybrid interaction system $\{\mathcal{G}_{t_k}\}_{k=1}^K$ for each dataset by averagely splitting the time interval of the dataset into $K$ time slots from the earliest click to the latest, defined by pivot timestamps: $\{t_k\}_{k=1}^K$. During the experiment, the number of intervals $K$ is searched from $\{2, 3, 4\}$.

\subsubsection{Compared Baselines}
\input{tables/main_result}

To demonstrate the effectiveness of our work, we compare \method{} with the current state-of-the-art baseline methods. Specifically, we choose the baseline methods from three different perspectives: (a) sequence-based recommendation(SR) models; (b) graph-based collaborative filtering methods; (c) continuous-time recommendation methods; \R{(d) contrastive learning-based recommendation models}.

% The baseline methods to be compared are as follows. \textbf{(a)} GRU4Rec \cite{hidasi2015session}: A SR model that leverages Recurrent Neural Networks (RNNs) to make predictions. \textbf{(a)} SASRec \cite{kang2018self}: A transformer-based SR model that introduces attention mechanism into recommendations. \textbf{(a)} TiSASRec \cite{li2020time}: A variant of SASRec that takes time intervals between interactions into consideration. It is one of the state-of-the-art attention-based SR baselines. \textbf{(b)} SR-GNN \cite{wu2019session}: A gated graph neural network (GGNN)-based session recommendation method. \textbf{(b)} LightGCN \cite{he2020lightgcn}: A classic GNN-based collaborative filtering method that uses GCNs in recommendation tasks. \textbf{(b)} DGCF \cite{wang2020disentangled}: It is a variant of LightGCN, which introduces disentangled representation learning into collaborative filtering. It's one of the state-of-the-art graph recommendation models. \textbf{(c)} TGSRec \cite{fan2021continuous}: A dynamic temporal graph-based model that integrate transformer with graph recommendations. \textbf{(c)} NODE \cite{bao2021time}: A neural ordinary differential equation (NODE)-based recommendation method that combines neural ODE with LSTMs to make sequence-based recommendations.
% \smallskip\noindent\textbf{(a) Sequence-based models: }
\begin{itemize}[leftmargin=*]
    \item \textbf{(a)} GRU4Rec \cite{hidasi2015session}: A SR model that leverages Recurrent Neural Networks (RNNs) to make predictions.
    \item \textbf{(a)} SASRec \cite{kang2018self}: A transformer-based SR model that introduces attention mechanism into recommendations.
    \item \textbf{(a)} TiSASRec \cite{li2020time}: A variant of SASRec that takes time intervals between interactions into consideration. It is one of the state-of-the-art attention-based SR baselines.
    \item \textbf{(b)} SR-GNN \cite{wu2019session}: A gated graph neural network (GGNN)-based session recommendation method.
    \item \textbf{(b)} LightGCN \cite{he2020lightgcn}: A classic GNN-based collaborative filtering method that uses GCNs in recommendation tasks.
    \item \textbf{(b)} DGCF \cite{wang2020disentangled}: It is a variant of LightGCN, which introduces disentangled representation learning into collaborative filtering. It's one of the state-of-the-art graph recommendation models.
    \item \textbf{(c)} TGSRec \cite{fan2021continuous}: A dynamic temporal graph-based model that integrates transformer with graph recommendations.
    \item \textbf{(c)} NODE \cite{bao2021time}: A neural ordinary differential equation (NODE)-based recommendation method that combines neural ODE with LSTMs to make sequence-based recommendations.
    \item \R{\textbf{(c)} GNG-ODE \cite{guo2022evolutionary}: A graph-based model that depicts the continuous-time session graphs with neural ODEs.}
    \item \R{\textbf{(d)}} \R{IOCRec} \cite{li2023multi}: \R{A sequence based contrastive learning approach that focuses on the intent-level self-supervision signals to promote model performance.}
    \item \R{\textbf{(d)}} \R{DCRec} \cite{yang2023debiased}: \R{A contrastive-based model that unifies sequential pattern encoding with global collaborative relation modeling in sequence recommendation.}
\end{itemize}

\subsubsection{Model Settings}
When evaluating our \method{} as well as all the compared baselines, we fix the embedding size for the user and items as 64 and the embedding size for the trainable time encoding vector as 16 if used. All models are optimized with the learning rate fixed $lr=0.001$. For a fair comparison, we utilize the same L2 normalization with fixed $\lambda=10^{-3}$. The max sequence length for sequence-based models is fixed at 50. Specifically, the step size of the \emph{Runge-Kutta} method for ODE-based models (\method{}, NODE) is fixed as $0.2$. \R{Specifically, for interaction graph-based methods (LightGCN, DGCF and TGSRec), we obtain the user-item graph from the interactions within the train set, and the number graph propagation layers is searched from $\{2,3,4\}$. The number of disentanglement of DGCF is searched from $\{2,4,8\}$. For TGSRec, we sample $10$ temporal neighbors at each layer of temporal graph.} % Moreover, we use the early stopping strategy to control the training process with a patience of 10, which means that the training will be stopped after 10 successive epochs without improvement on the validation set.
All of the mentioned methods are implemented using PyTorch and torchdiffeq\footnote{{\url{https://github.com/rtqichen/torchdiffeq}}}.% The baseline methods are based on the open-sourced code or implemented according to the original paper.

To evaluate the model performance, we adopt Recall@K and MRR as the evaluation metrics, following the advice of previous works \cite{fan2021continuous,bao2021time}. To be specific, we calculate Recall@5, Recall@10 and MRR along with all the items ranked by the model. The model that achieves the highest MRR on the evaluation set is selected to be tested on the test set and the test results are shown in table \ref{tab:overall}.

\subsubsection{Experimental Results}

From the results reported in table \ref{tab:overall}, we can make the following observations:
\begin{itemize}[leftmargin=*]
    \item User-item graph-based baseline methods (LightGCN and DGCF) outperform sequence-based methods when the datasets are rather sparse (i.e. Cloth, and Music), where the high-order similarities between users and items can provide useful information. On the other hand, sequence-based models (GRU4Rec, SASRec, and TiSASRec) yield better results on dense datasets. \R{Specifically, contrastive learning-based method (IOCRec and DCRec) shows more competitive performance, which demonstrates the effectiveness of contrastively learning to represent items.}% (i.e. ML-1m and ML-100K). which shows the importance of explicitly modeling the temporal influence on the contextual information for users with long history sequence.
    \item ODE-based methods (\R{NODE, GNG-ODE and \method{}}) generally achieve promising performance on both sparse and dense datasets, which shows the effectiveness of depicting the continuous dynamics of user behaviors. Neural ODE helps to overcome the difficulties brought by irregularly sampled interactions, as well as to capture the implicit dynamics in long-term interacting systems.% capture the evolving process of user interest and 
    \item \method{} outperforms all baseline methods on all datasets. In particular, \method{} has gained more than $6.5\%$ and $8.8\%$ relative improvement at Recall@5 and Recall@10, as well as a $2\%$ relative improvement at MRR. The improvement made by \method{} benefited from building a hybrid dynamic interaction system and a well-designed graph propagation scheme.
\end{itemize}

\subsection{Ablation and Parameter Studies (RQ2)}
In this section, we further investigate the detailed functionality of different modules in \method{}, as well as their performance under various kinds of input temporal signals by conducting ablation studies. We will also explore \method{}'s sensitivity to the hyper-parameters.% via a series of parameter studies.

\subsubsection{Functionality of Graph Modules}
\input{tables/module_ablation}

\begin{figure*}[t]
\centering
\includegraphics[width=0.95\linewidth]{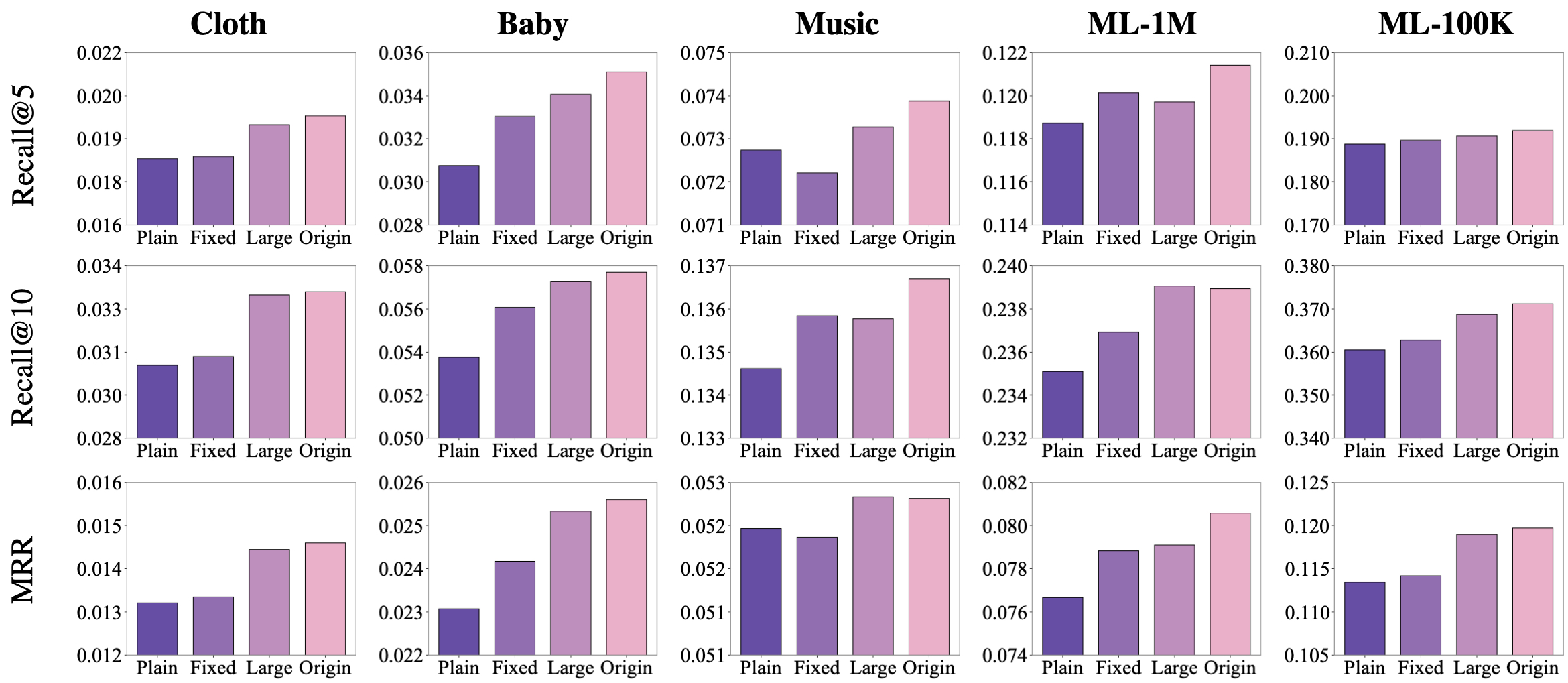}
% \begin{subfigure}{0.85\linewidth}
% \caption{Results on \textbf{Music}.}
% \begin{subfigure}{0.32\linewidth}
%     \includegraphics[width=\linewidth]{ figures/temporal_enc/music_recall_5.pdf}
% \end{subfigure}
% \begin{subfigure}{0.32\linewidth}
%     \includegraphics[width=\linewidth]{ figures/temporal_enc/music_recall_10.pdf}
% \end{subfigure}
% \begin{subfigure}{0.32\linewidth}
%     \includegraphics[width=\linewidth]{ figures/temporal_enc/music_mrr.pdf}
% \end{subfigure}
% \end{subfigure}

% \begin{subfigure}{0.85\linewidth}
% \caption{Results on \textbf{ML-1M}.}
% \begin{subfigure}{0.32\linewidth}
%     \includegraphics[width=\linewidth]{ figures/temporal_enc/ML_recall_5.pdf}
% \end{subfigure}
% \begin{subfigure}{0.32\linewidth}
%     \includegraphics[width=\linewidth]{ figures/temporal_enc/ML_recall_10.pdf}
% \end{subfigure}
% \begin{subfigure}{0.32\linewidth}
%     \includegraphics[width=\linewidth]{ figures/temporal_enc/ML_mrr.pdf}
% \end{subfigure}
% \end{subfigure}
\caption{Performance comparison w.r.t. different types of temporal encoding functions.}
\label{fig:abli_temporal_enc}
\end{figure*}
Since we have proposed two different graph propagation schemes: an ODE-based edge evolving module and an attention-based temporal aggregating module, it's necessary to conduct ablation studies to demonstrate the effectiveness of the two tailored-designed propagation modules. To be specific, we consider the following variants of \method{} to explore how both modules cooperate for better recommendation results.

To start with, we test the model performance by removing the temporal attention module and replacing the module with the plain Graph Convolution Networks (GCNs) respectively to verify the effectiveness of the temporal aggregating module, namely \method{}\textsubscript{ODE} and \method{}\textsubscript{GCN}. Then we remove the edge evolving module from the original model, namely \method{}\textsubscript{Att}. % We choose two representative datasets, Music and ML-1M to conduct the experiments. 
From the results in Table \ref{tab:ablation} we can observe that:
\begin{itemize}[leftmargin=*]
    \item Both the edge evolving module and temporal aggregating module are important for \method{} to achieve better performance. Specifically, \method{}\textsubscript{ATT} suffers the most performance decline for removing the ODE layers from the original architecture.% The result implies that compared with aggregating neighboring information, depicting the evolving dynamics in the interaction system is more fundamental to \method{}.
    %\item Compared with the original model, \method{}\textsubscript{ODE}, which deletes the attention-based temporal aggregating module, also suffers from a performance decline. The result shows that the attention module boosts the recommendation effect by dynamically modeling the relativity and high-order connectivity between user and item nodes during time intervals.
    \item The comparison between \method{} and \method{}\textsubscript{GCN} demonstrates the superiority of our proposed temporal attention module over regular graph convolutions. \method{} outperforms more over \method{}\textsubscript{GCN} on datasets that include longer time spans (i.e. Cloth, Music), rather than datasets within a relatively short time period (ML-1M). The result reveals the strength of \method{} in dealing with long-term interacting systems.%effectiveness of our proposed temporal attention aggregating module. The temporal attention mechanism boosts the model performance by explicitly encoding the temporal factor and the affinity between users and items.
\end{itemize}

\subsubsection{Influence of Temporal Encoding}

In the attention module, a set of learnable time encoding functions have been defined in Eq. \ref{eq:temporal_encode} to bring more temporal information to the module. To validate the effectiveness of the proposed encoding functions, we conduct a series of ablation studies by replacing the attention module with (a) a plain attention module without temporal encodings, (b) temporal encodings with fixed random weights, (c) default temporal encodings with $d_T=16$, and (d) extreme large temporal encodings with $d_T=64$. As illustrated in Figure \ref{fig:abli_temporal_enc}, we observe that:
\begin{itemize}[leftmargin=*]
    \item Time encodings are necessary for \method{} to better leverage temporal information. In general, the plain attention method without time encodings would lead to sub-optimal results.
    \item In most cases, learnable time encodings are better than fixed parameters for self-adapting to flexible time spans. However, extremely large time vectors could lead noises into attention modules and cause a performance decline.
\end{itemize}

\subsubsection{Influence of System Dynamic}

\begin{figure*}
\centering
% Results on \textbf{Music}.
% \begin{subfigure}{0.85\linewidth}
%     \caption{Results on \textbf{Music}.}
% \begin{subfigure}{0.32\linewidth}
%     \includegraphics[width=\linewidth]{ figures/abli_K/music_recall_5.pdf}
% \end{subfigure}
% \begin{subfigure}{0.32\linewidth}
%     \includegraphics[width=\linewidth]{ figures/abli_K/music_recall_10.pdf}
% \end{subfigure}
% \begin{subfigure}{0.32\linewidth}
%     \includegraphics[width=\linewidth]{ figures/abli_K/music_mrr.pdf}
% \end{subfigure}
% \end{subfigure}

% % Results on \textbf{ML-1M}.
% \begin{subfigure}{0.85\linewidth}
%     \caption{Results on \textbf{ML-1M}.}
% \begin{subfigure}{0.32\linewidth}
%     \includegraphics[width=\linewidth]{ figures/abli_K/ML_recall_5.pdf}
% \end{subfigure}
% \begin{subfigure}{0.32\linewidth}
%     \includegraphics[width=\linewidth]{ figures/abli_K/ML_recall_10.pdf}
% \end{subfigure}
% \begin{subfigure}{0.32\linewidth}
%     \includegraphics[width=\linewidth]{ figures/abli_K/ML_mrr.pdf}
% \end{subfigure}
% \end{subfigure}

\includegraphics[width=0.95\linewidth]{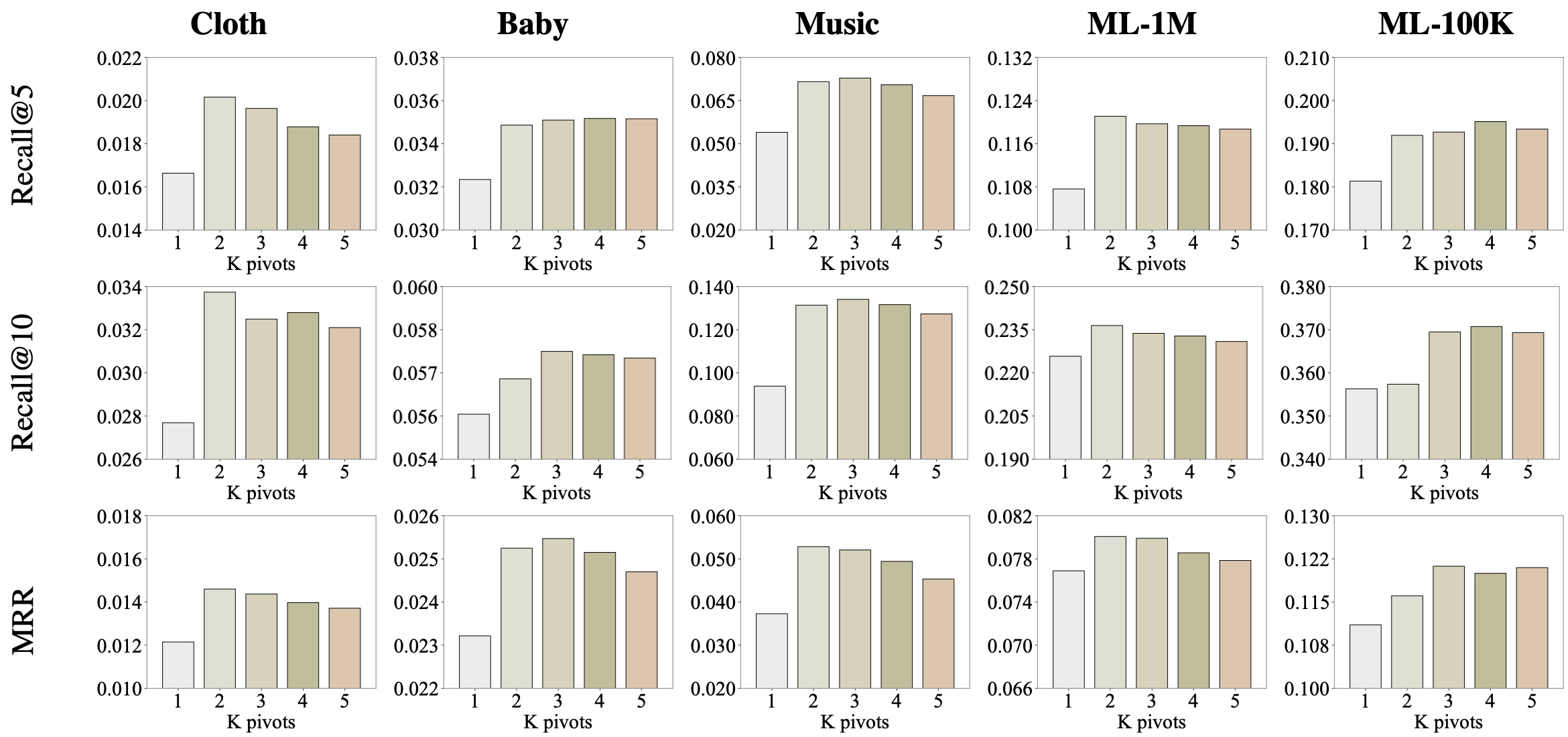}
\caption{Performance comparison w.r.t. different settings of the number of pivot timestamps $K$.}
\label{fig:abli_K}
\end{figure*}
\input{tables/signal_ablation}

Recalling that we build a hybrid dynamic interaction system $\{\mathcal{G}_{t_k}\}_{k=1}^K$ based on the time interval of the original interaction dataset, thus the system dynamic is influenced by the defined pivot timestamps $\{t_k\}_{k=1}^K$. To be specific, by increasing $K$, we are adding more pivot timestamps to the system and thus introducing more detailed dynamic features for \method{} in the ODE-based edge evolving process. From the results illustrated in Figure \ref{fig:abli_K}, we can observe that:

\begin{itemize}[leftmargin=*]
    \item By setting a larger $K$, the model could benefit from the detailed dynamics of the interacting system. Generally, the performance of \method{} reaches the optimal when $K$ is set by 3 as a trade-off between performance and computational cost.
    \item Too many time intervals (more than 4) may cause a decline in the model performance on both datasets.% It's caused by the over-smoothing issue of the GNNs applied in the temporal aggregating module. Thus to extend the depth of the model, we should not increase the depth of the aggregating module but the depth of the ODE-based edge evolving module.
\end{itemize}

\subsubsection{Model's Sensitivity to Temporal Signals}

In \method{}, the interactions between users and items are split into a set of hybrid dynamic systems $\{\mathcal{G}_{t_k}\}_{k=1}^K$. By default, the two components of the autoregressive framework defined in Eq. \ref{eq:regressive} receive interactions with different temporal signals. Specifically, the edge evolving module is concerned about the interactions within just the current time interval, while the temporal aggregating module would aggregate the neighbors that belong to current or previous intervals.

\begin{figure*}[t]
\centering
\includegraphics[width=0.95\linewidth]{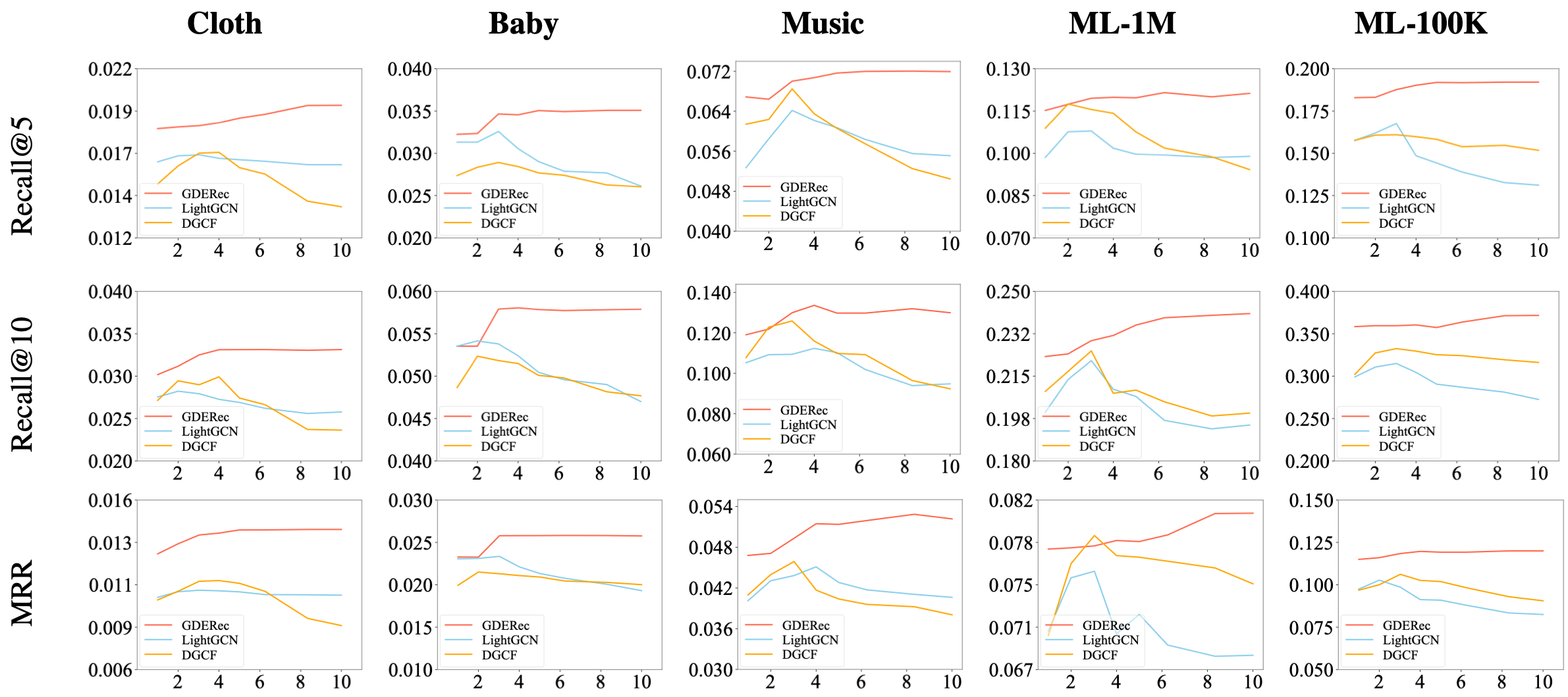}
% \begin{subfigure}{0.85\linewidth}
% \caption{Results on \textbf{Music}.}
% \begin{subfigure}{0.32\linewidth}
%     \includegraphics[width=\linewidth]{ figures/ode_step/step_music_recall_5.pdf}
% \end{subfigure}
% \begin{subfigure}{0.32\linewidth}
%     \includegraphics[width=\linewidth]{ figures/ode_step/step_music_recall_10.pdf}
% \end{subfigure}
% \begin{subfigure}{0.32\linewidth}
%     \includegraphics[width=\linewidth]{ figures/ode_step/step_music_mrr.pdf}
% \end{subfigure}
% \end{subfigure}

% \begin{subfigure}{0.85\linewidth}
% \caption{Results on \textbf{ML-1M}.}
% \begin{subfigure}{0.32\linewidth}
%     \includegraphics[width=\linewidth]{ figures/ode_step/step_ml_recall_5.pdf}
% \end{subfigure}
% \begin{subfigure}{0.32\linewidth}
%     \includegraphics[width=\linewidth]{ figures/ode_step/step_ml_recall_10.pdf}
% \end{subfigure}
% \begin{subfigure}{0.32\linewidth}
%     \includegraphics[width=\linewidth]{ figures/ode_step/step_ml_mrr.pdf}
% \end{subfigure}
% \end{subfigure}
\caption{Performance comparison w.r.t. model depth. For our \method{}, we define the depth of the edge evolving module by the number of steps the ODE solver takes to calculate the numerical solution. In other words, the little the step size is, the deeper \method{} is.}
\label{fig:ode_step}
\end{figure*}
However, we wonder if this set of input temporal signals can fully exploit the potential of \method{}'s autoregressive framework to better leverage the dynamics in an interacting system. Thus we conduct experiments on \method{} with different combinations of input signals, listed as follows:
\begin{enumerate}[leftmargin=*]
    \item \textit{$M_1$. Static System}: The ODE module and the attention module would both receive all observed interactions as input.% This setting completely breaks the original dynamic characteristics of interacting systems and reflects \method{}'s capability of dealing with static input situations.
    \item \textit{$M_2$. Short-sighted Model}: Both modules are limited to using only the interactions within the current interval.% Under this setting the temporal aggregation module would suffer from its short-sighted view with limited temporal neighbors.
    \item \textit{$M_3$. Reversed Sights}: In contrast to the original setting, the ODE module can take previous interactions as input and the attention module is limited to only the current interval.
    \item \textit{$M_4$. Look-back Model}: Both modules now can leverage interactions from history so far.% The ODE module is overloaded with numerous interactions under this setting.
\end{enumerate}

The experimental results are illustrated in table \ref{tab:temporal_signal} of all the mentioned methods, comparing with the original modal. We can observe from the results that:
\begin{itemize}[leftmargin=*]
    \item In most cases, compared with static interactions as inputs ($M_1$), inputs with different degrees of dynamic signals ($M_2$-$M_4$) show better performance, which demonstrates that it is necessary to introduce dynamic temporal signals.
    \item Compared with other combinations, the original input of \method{} stays stable advantage on all datasets. On the one hand, the ODE module would receive only the latest interactions and thus avoid the overloaded information brought by previous long sequences and depict the fine-grained interaction dynamics. On the other hand, the attention module could leverage all historical information to adaptively respond to the growing interaction graph.

\end{itemize}

\subsubsection{Influence of Model Depth}

As one of the advantages of neural ODEs, we can extend the model depth by controlling the \emph{number of function evaluations} (NFE) of the numerical ODE solver, without concerns about the over-smoothing issue brought by GNN layers. In particular, we vary the model depth from 1 (discrete form) to 10 with fixed $K=2$ to validate the influence of the depth of the edge evolving module. We conduct experiments on the Music dataset and from the results shown in Figure \ref{fig:ode_step} we can observe that:
\begin{itemize}[leftmargin=*]
    \item With the increase in model depth, GNN-based models (LightGCN and DGCF) will suffer from the over-smoothing issue and the decline in performance. On the other hand, ODE-based \method{} benefits from increasing the model depth, indicating the capability of Neural ODEs to capture the long-term evolving features.
    \item When the model becomes deeper, the performance of \method{} will firstly reach the optimal and then maintain stable performance, showing its robustness to exploit dynamic interactions. However, the growth of model performance slows down when the model is too deep (more than 6 layers), where stacking more layers could gain little performance growth.
\end{itemize}

\begin{figure}[t]
\centering
\begin{subfigure}{\linewidth}
    \includegraphics[width=\linewidth]{ 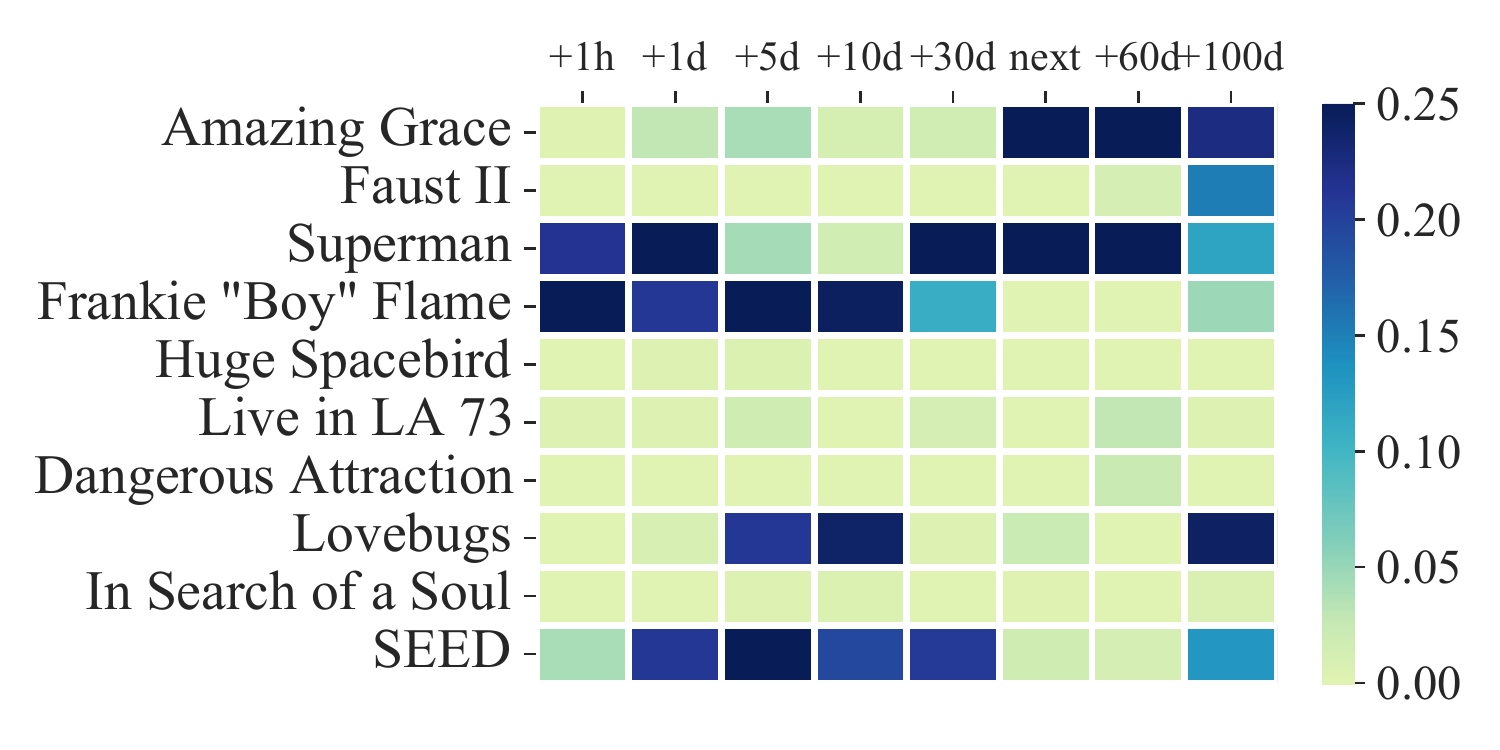}
\end{subfigure}
\caption{Visualization of attention Weights.}
\label{fig:heat}
\end{figure}
\subsection{Visualization and Case Study (RQ3)}
\begin{figure*}[t]
\centering
\begin{subfigure}{\linewidth}
    \includegraphics[width=\linewidth]{ 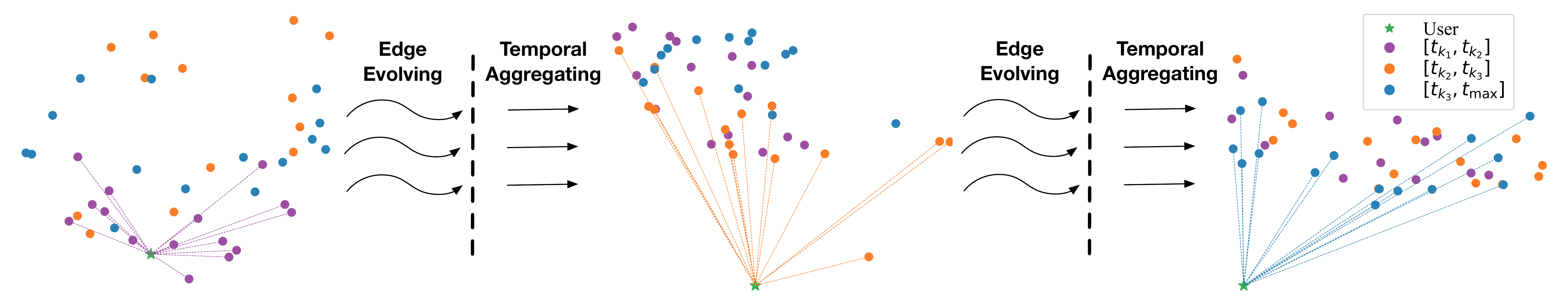}
\end{subfigure}
\caption{Visualization of model output.}
\label{fig:t-sne}
\end{figure*}

\begin{figure*}[t]
\centering
\begin{subfigure}{\linewidth}
    \includegraphics[width=\linewidth]{ 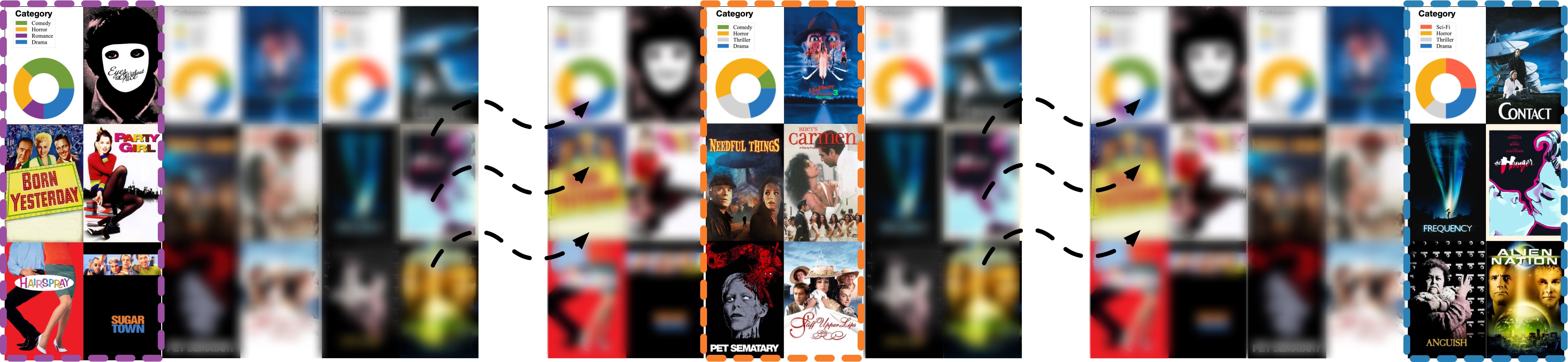}
\end{subfigure}
\caption{Case study of the selected user $u$.}
\label{fig:case}
\end{figure*}

Since we have proposed to model the dynamic factors to model an irregularly-sampled interaction system, we attempt to understand how users and items evolve as \method{} propagates. Towards this end, we randomly choose a user $u$ from the ML-1M dataset and visualize the hidden representation of $u$ as well as the clicked movies during different periods and conduct a series of studies to investigate how \method{} alleviate the irregularly-sampled issue.

\subsubsection{Visualization of Temporal Attention}

The use of temporal encodings in the attention module brings fine-grained temporal information, which helps the model respond to different time conditions. Specifically, we choose a random user from the amazon-music dataset and sampled 10 songs from his historical clicks. We obtain the output logits on these songs of the attention module, followed by a soft-max normalization, and visualize them in a heat map in Figure \ref{fig:heat}. Each row represents a time increment since the first interaction of the user, and row "next" represents the day when the test set begins. From the figure, we can observe the dynamic evolution of user interest and the temporal encodings empower the attention module to capture the fluid interests.

\subsubsection{Case Study on Hybrid Dynamic System}
As the hidden representations evolve as time proceeds, we attempt to visualize the influence of the temporal factors in the node embedding space. To be specific, we visualize the hidden representation of $u$ and the movies he/she has clicked on each period $[t_k,t_{k+1}],1\leq k\leq K$ to illustrate how the embedding space of irregularly-sampled items evolves with the dynamic interaction system. We use the t-SNE algorithm \cite{van2008visualizing} to visualize the representations of movies being clicked in different periods, which are marked with corresponding colors, as shown in Figure \ref{fig:t-sne}. 

As illustrated in Figure \ref{fig:t-sne}, we can observe the cluster structure of the interacted items during different time periods. The distance between the user $u$ (marked with green stars) and items (marked with corresponding colored dots) changes over time periods, reflecting the effects of the edge evolving module. The interactions generate attractive forces between users and items and the neural ODE process depicts this evolution effect in the embedding space. The visualization illustrates \method{} can encode items sampled at irregular timestamps and depict the dynamic evolution in the latent space.

We further show parts of the movies $u$ has rated\footnote{{The movie posters are downloaded from MovieLens website \url{https://movielens.org}}.} during different time periods in Figure \ref{fig:case}. By arranging the rated movies in temporal order, we can observe the favorable change of $u$ in different movie categories. During the first period $u$ preferred comedies and dramas, before his/her taste gradually changed to horror films and thrillers in the second period, and during the last period of time, $u$  is shown to develop new interests in Sci-Fis. Compared with the visualization in Figure \ref{fig:t-sne}, we can observe that \method{} has successfully learned the temporal preference of $u$, and the evolving process is reflected by the item embeddings.

%% file: tables/main_result.tex
\begin{table*}
\centering
\caption{The test results of \method{} and all baselines on the five real-world datasets, where R@K is short for Recall@K. The highest performance is emphasized with bold font and the second highest is marked with underlines.\textbf{$\star$} indicates that \method{} outperforms the best baseline model at a p-value$<$0.05 level of unpaired t-test.}
\label{tab:overall}
\renewcommand{\arraystretch}{1.1}
\begin{tabularx}{\linewidth}{cYYYYYYYYYYYYYYY}
\toprule %\hline
\multirow{2}{*}{Model} & \multicolumn{3}{c}{Cloth} & \multicolumn{3}{c}{Baby} & \multicolumn{3}{c}{Music} & \multicolumn{3}{c}{ML-1M} & \multicolumn{3}{c}{ML-100K}\\ 
\cmidrule[0.5pt](lr){2-4}\cmidrule[0.5pt](lr){5-7}\cmidrule[0.5pt](lr){8-10}\cmidrule[0.5pt](lr){11-13}\cmidrule[0.5pt](lr){14-16}
& R@5$\uparrow$ & R@10$\uparrow$ & MRR$\uparrow$ & R@5 & R@10 & MRR & R@5 & R@10 & MRR & R@5 & R@10 & MRR & R@5 & R@10 & MRR\\
\midrule 
GRU4Rec & 0.0106 & 0.0212 & 0.0090
& 0.0295 & 0.0521 & 0.0219
& 0.0375 & 0.0544 & 0.0216
& 0.1065 & 0.2131 & 0.0755
& 0.1769 & 0.3475 & 0.1136 \\
SASRec & 0.0115 & 0.0206 & 0.0099
& 0.0309 & 0.0548 & 0.0215
& 0.0429 & 0.0771 & 0.0320
& 0.1010 & 0.2135 & 0.0787
& 0.1675 & 0.3064 & 0.1050 \\
TiSASRec & 0.0112 & 0.0214 & 0.0106
& 0.0318 & {0.0557} & {0.0234}
& 0.0510 & 0.0825 & 0.0322
& 0.1121 & 0.2253 & 0.0778
& {0.1792} & {0.3552} & 0.1108 \\
\midrule
SR-GNN & 0.0058 & 0.0103 & 0.0045
& 0.0140 & 0.0276 & 0.0115
& 0.0359 & 0.0650 & 0.0295
& 0.1091 & 0.2187 & 0.0744
& 0.1781 & 0.3422 & 0.1011 \\
LightGCN & 0.0169 & 0.0283 & 0.0106
& 0.0325 & 0.0537 & 0.0233
& 0.0642 & 0.1167 & 0.0451
& 0.1079 & 0.2214 & 0.0757
& 0.1676 & 0.3128 & 0.1028 \\
DGCF & 0.0170 & 0.0289 & 0.0112
& 0.0289 & 0.0518 & 0.0213
& \underline{0.0695} & \underline{0.1258} & \underline{0.0459}
& {0.1151} & 0.2276 & {0.0789}
& 0.1601 & 0.3329 & 0.1062 \\
\midrule
\R{IOCRec} & {0.0129} & {0.0264} & {0.0115} & 0.0329 & 0.0542 & 0.0229 & 0.0649 & 0.1178 & 0.0440 & 0.1149 & 0.2259 & 0.0772 & \underline{0.1855} & \underline{0.3622} & \underline{0.1175} \\
\R{DCRec} & 0.0117 & 0.0251 & 0.0108 & 0.0335 & 0.0551 & 0.0227 & 0.0591 & 0.1062 & 0.0433 & \underline{0.1160} & {0.2285} & \underline{0.0791} & 0.1802 & 0.3541 & 0.1162 \\
\midrule
TGSRec & 0.0104 & 0.0271 & 0.0114
& 0.0142 & 0.0240 & 0.0116
& 0.0243 & 0.0377 & 0.0158
& 0.1109 & 0.2123 & 0.0749
& 0.1601 & 0.3277 & {0.1120} \\
NODE & {0.0184} & {0.0293} & {0.0117}
& {0.0339} & 0.0525 & 0.0224
& 0.0584 & 0.0967 & {0.0420}
& 0.1148 & \underline{0.2297} & 0.0785
& 0.1739 & 0.3404 & 0.1098 \\
\R{GCG-ODE} & \underline{0.0186} & \underline{0.0299} & \underline{0.0123} &\underline{0.0340} & \underline{0.0558} & \underline{0.0238} & 0.0682 & 0.1237 & 0.0455 &  0.1149 & 0.2263 & 0.0767 & 0.1846 & 0.3598 & 0.1161 \\
\midrule
\method{} & \textbf{0.0198\textsuperscript{$\star$}} & \textbf{0.0331\textsuperscript{$\star$}} & \textbf{0.0146\textsuperscript{$\star$}} 
& \textbf{0.0351\textsuperscript{$\star$}} & \textbf{0.0577\textsuperscript{$\star$}} & \textbf{0.0256\textsuperscript{$\star$}} 
& \textbf{0.0738\textsuperscript{$\star$}} & \textbf{0.1363\textsuperscript{$\star$}} & \textbf{0.0528\textsuperscript{$\star$}} 
& \textbf{0.1210\textsuperscript{$\star$}} & \textbf{0.2389\textsuperscript{$\star$}} & \textbf{0.0801\textsuperscript{$\star$}} 
& \textbf{0.1919\textsuperscript{$\star$}} & \textbf{0.3712\textsuperscript{$\star$}} & \textbf{0.1197\textsuperscript{$\star$}} \\
\bottomrule % \hline
\end{tabularx}
\end{table*}

%% file: tables/module_ablation.tex
\begin{table*}
\centering
\caption{\method{}'s recommendation results with different settings of graph modules.}
\label{tab:ablation}
\renewcommand{\arraystretch}{1.1}
\setlength{\tabcolsep}{3pt}
\begin{tabularx}{\linewidth}{cYYYYYYYYYYYYYYY}
\toprule %\hline
\multirow{2}{*}{Dataset} & \multicolumn{3}{c}{Cloth} & \multicolumn{3}{c}{Baby} & \multicolumn{3}{c}{Music} & \multicolumn{3}{c}{ML-1M} & \multicolumn{3}{c}{ML-100K} \\
\cmidrule[0.5pt](lr){2-4}\cmidrule[0.5pt](lr){5-7}\cmidrule[0.5pt](lr){8-10}\cmidrule[0.5pt](lr){11-13}\cmidrule[0.5pt](lr){14-16}
& R@5$\uparrow$ & R@10$\uparrow$ & MRR$\uparrow$ & R@5 & R@10 & MRR  & R@5 & R@10 & MRR  & R@5 & R@10 & MRR  & R@5 & R@10 & MRR \\
\midrule
LightGCN & 0.0169 & 0.0283 & 0.0106
& 0.0325 & 0.0537 & 0.0233
& 0.0642 & 0.1167 & 0.0451
& 0.1079 & 0.2214 & 0.0757
& 0.1676 & 0.3128 & 0.1028 \\
\method{}\textsubscript{ATT} &  0.0160 & 0.0279 & 0.0113 & 0.0317 & 0.0533 & 0.0229 & 0.0641 & 0.1178 & 0.0461 & 0.1132 & 0.2211 & 0.0768 & 0.1909 & 0.3733 & 0.1138 \\
\method{}\textsubscript{ODE} & 0.0173 & 0.0295 & 0.0125 & 0.0323 & 0.0565 & 0.0245 & 0.0693 & 0.1280 & 0.0519 & 0.1061 & 0.2243 & 0.0746 & 0.1739 & 0.3571& 0.1095 \\
\method{}\textsubscript{GCN} & 0.0187 & 0.0303 & 0.0134 & 0.0349 & 0.0574 & \textbf{0.0260} & 0.0707 & 0.1291 & 0.0506 & 0.1192 & 0.2388 & 0.0799 & 0.1887 & 0.3690 & 0.1101 \\
\method{} & \textbf{0.0198} & \textbf{0.0331} & \textbf{0.0146} 
& \textbf{0.0351} & \textbf{0.0577} & 0.0256
& \textbf{0.0738} & \textbf{0.1363} & \textbf{0.0528} 
& \textbf{0.1210} & \textbf{0.2389} & \textbf{0.0801} 
& \textbf{0.1919} & \textbf{0.3712} & \textbf{0.1197} \\
\bottomrule %\hline
\end{tabularx}
\end{table*}

%% file: tables/signal_ablation.tex
\begin{table*}[t] 
\footnotesize
\caption{Results of ablation studies on \method{}'s sensitivity towards input temporal signals. Cur. means that each corresponding module at $k$-th layer could only observe interactions within the current interval ($[t_{k}, t_{k+1}]$). Prev. means the previous ($[t_0, t_{k+1}]$) interactions are visible, and All means all interactions are visible during the training and inference process.
\label{tab:temporal_signal}}
\renewcommand{\arraystretch}{1.1}
\centering
\resizebox{\textwidth}{!}{
    \begin{tabularx}{\linewidth}{cYYYYYYYYYYYYYY}
    \toprule
    \multirow{2}{*}{Method} & \multicolumn{2}{c}{Input Signal} & \multicolumn{3}{c}{Cloth} & \multicolumn{3}{c}{Baby} & \multicolumn{3}{c}{Music} & \multicolumn{3}{c}{ML-1M}\\% & \multicolumn{3}{c}{ML-100K} \\ 
    \cmidrule[0.5pt](lr){4-6}\cmidrule[0.5pt](lr){7-9}\cmidrule[0.5pt](lr){10-12}\cmidrule[0.5pt](lr){13-15}%\cmidrule[0.5pt](lr){16-18}
    & ODE & Attn. & R@5$\uparrow$ & R@10$\uparrow$ & MRR$\uparrow$ & R@5 & R@10 & MRR & R@5 & R@10 & MRR & R@5 & R@10 & MRR \\%& R@5 & R@10 & MRR \\
    \midrule
    $M_1$ & All & All & 0.0181 & 0.0321 & 0.0133 & 0.0338 & 0.0523 & 0.0238 & 0.0687 & 0.1233 & 0.0502 & 0.1108 & 0.2217 & 0.0769 \\%& 0.1853 &  0.3584 & 0.1146 \\
    $M_2$ & Cur. & Cur. & 0.0185 & 0.0332 & 0.0137 & 0.0336 & 0.0570 & 0.0253 & \textbf{0.0728} & 0.1296 & 0.0511 & 0.1251 & 0.2280 & 0.0783 \\% & 0.1803 & 0.3637 & 0.1121\\
    $M_3$ & Prev. & Cur. &  0.0182 & \textbf{0.0336} & 0.0132 & 0.0330 & 0.0549 & 0.0241 & 0.0679 & 0.0128 & 0.0494 & 0.1201 & 0.2377 & 0.0784 \\% & 0.1835 & 0.3616 & 0.1186 \\
    $M_4$ & Prev. & Prev. & 0.1783 & 0.0309 & 0.0126 & 0.0312 & 0.0538 & 0.0229 & 0.0715 & 0.1278 & 0.0517 & 0.1207 & \textbf{0.2397} & 0.0783 \\% & 0.1803 & 0.3574 & \textbf{0.1212} \\
    Origin & Cur. & Prev. & \textbf{0.0198} & 0.0331 & \textbf{0.0146} & \textbf{0.0351} & \textbf{0.0577} & \textbf{0.0256} & 0.0738 & \textbf{0.1363} & \textbf{0.0528} & \textbf{0.1210} & 0.2389 & \textbf{0.0801} \\% & \textbf{0.1919} & \textbf{0.3712} & 0.1197 \\
    \bottomrule
    \end{tabularx}
}
\end{table*}

%% file: 6_conclusion.tex
\section{Conclusion}

In this work, we propose an autoregressive ODE-based graph recommendation framework \method{}, for graph recommendation on the hybrid dynamic interacting system, which is built upon two tailored designed graph propagation modules. On the one hand, a neural ODE-based edge evolving module implicitly depicts the dynamic evolving process brought by the collaborative affinity between user-item interactions. On the other hand, a temporal attention-based graph aggregating module explicitly aggregates neighboring information on the interaction graph. We define the way to build the corresponding hybrid dynamic interaction systems given a temporal interaction graph $\mathcal{G}$. By autoregressively applying the two modules, \method{} obtains node representations with temporal and dynamic features on a hybrid dynamic system. Comprehensive experiments and visualization results on several real-world datasets illustrate the effectiveness and strength of \method{}.

%% file: main.bbl
\begin{thebibliography}{10}
\providecommand{\url}[1]{#1}
\csname url@samestyle\endcsname
\providecommand{\newblock}{\relax}
\providecommand{\bibinfo}[2]{#2}
\providecommand{\BIBentrySTDinterwordspacing}{\spaceskip=0pt\relax}
\providecommand{\BIBentryALTinterwordstretchfactor}{4}
\providecommand{\BIBentryALTinterwordspacing}{\spaceskip=\fontdimen2\font plus
\BIBentryALTinterwordstretchfactor\fontdimen3\font minus
  \fontdimen4\font\relax}
\providecommand{\BIBforeignlanguage}[2]{{%
\expandafter\ifx\csname l@#1\endcsname\relax
\typeout{** WARNING: IEEEtran.bst: No hyphenation pattern has been}%
\typeout{** loaded for the language `#1'. Using the pattern for}%
\typeout{** the default language instead.}%
\else
\language=\csname l@#1\endcsname
\fi
#2}}
\providecommand{\BIBdecl}{\relax}
\BIBdecl

\bibitem{wang2018billion}
J.~Wang, P.~Huang, H.~Zhao, Z.~Zhang, B.~Zhao, and D.~L. Lee, ``Billion-scale
  commodity embedding for e-commerce recommendation in alibaba,'' in
  \emph{Proceedings of the 24th ACM SIGKDD International Conference on
  Knowledge Discovery \& Data Mining}, 2018, pp. 839--848.

\bibitem{li2020hierarchical}
Z.~Li, X.~Shen, Y.~Jiao, X.~Pan, P.~Zou, X.~Meng, C.~Yao, and J.~Bu,
  ``Hierarchical bipartite graph neural networks: Towards large-scale
  e-commerce applications,'' in \emph{2020 IEEE 36th International Conference
  on Data Engineering (ICDE)}.\hskip 1em plus 0.5em minus 0.4em\relax IEEE,
  2020, pp. 1677--1688.

\bibitem{zhang2022oa}
X.~Zhang, C.~Zhang, X.~Li, X.~L. Dong, J.~Shang, C.~Faloutsos, and J.~Han,
  ``Oa-mine: Open-world attribute mining for e-commerce products with weak
  supervision,'' in \emph{Proceedings of the ACM Web Conference 2022}, 2022,
  pp. 3153--3161.

\bibitem{min2022divide}
E.~Min, Y.~Rong, Y.~Bian, T.~Xu, P.~Zhao, J.~Huang, and S.~Ananiadou,
  ``Divide-and-conquer: Post-user interaction network for fake news detection
  on social media,'' in \emph{Proceedings of the ACM Web Conference 2022},
  2022, pp. 1148--1158.

\bibitem{mousavi2022effective}
M.~Mousavi, H.~Davulcu, M.~Ahmadi, R.~Axelrod, R.~Davis, and S.~Atran,
  ``Effective messaging on social media: What makes online content go viral?''
  in \emph{Proceedings of the ACM Web Conference 2022}, 2022, pp. 2957--2966.

\bibitem{song2022have}
J.~Song, K.~Han, and S.-W. Kim, ``“i have no text in my post”: Using visual
  hints to model user emotions in social media,'' in \emph{Proceedings of the
  ACM Web Conference 2022}, 2022, pp. 2888--2896.

\bibitem{mnih2007probabilistic}
A.~Mnih and R.~R. Salakhutdinov, ``Probabilistic matrix factorization,''
  \emph{Advances in neural information processing systems}, vol.~20, 2007.

\bibitem{gopalan2015scalable}
P.~Gopalan, J.~M. Hofman, and D.~M. Blei, ``Scalable recommendation with
  hierarchical poisson factorization.'' in \emph{UAI}, 2015, pp. 326--335.

\bibitem{koren2009matrix}
Y.~Koren, R.~Bell, and C.~Volinsky, ``Matrix factorization techniques for
  recommender systems,'' \emph{Computer}, vol.~42, no.~8, pp. 30--37, 2009.

\bibitem{rendle2010factorizing}
S.~Rendle, C.~Freudenthaler, and L.~Schmidt-Thieme, ``Factorizing personalized
  markov chains for next-basket recommendation,'' in \emph{Proceedings of the
  19th international conference on World wide web}, 2010, pp. 811--820.

\bibitem{he2016fusing}
R.~He and J.~McAuley, ``Fusing similarity models with markov chains for sparse
  sequential recommendation,'' in \emph{2016 IEEE 16th international conference
  on data mining (ICDM)}.\hskip 1em plus 0.5em minus 0.4em\relax IEEE, 2016,
  pp. 191--200.

\bibitem{hochreiter1997long}
S.~Hochreiter and J.~Schmidhuber, ``Long short-term memory,'' \emph{Neural
  computation}, vol.~9, no.~8, pp. 1735--1780, 1997.

\bibitem{hidasi2015session}
B.~Hidasi, A.~Karatzoglou, L.~Baltrunas, and D.~Tikk, ``Session-based
  recommendations with recurrent neural networks,'' \emph{arXiv preprint
  arXiv:1511.06939}, 2015.

\bibitem{quadrana2017personalizing}
M.~Quadrana, A.~Karatzoglou, B.~Hidasi, and P.~Cremonesi, ``Personalizing
  session-based recommendations with hierarchical recurrent neural networks,''
  in \emph{proceedings of the Eleventh ACM Conference on Recommender Systems},
  2017, pp. 130--137.

\bibitem{li2018learning}
Z.~Li, H.~Zhao, Q.~Liu, Z.~Huang, T.~Mei, and E.~Chen, ``Learning from history
  and present: Next-item recommendation via discriminatively exploiting user
  behaviors,'' in \emph{Proceedings of the 24th ACM SIGKDD International
  Conference on Knowledge Discovery \& Data Mining}, 2018, pp. 1734--1743.

\bibitem{vaswani2017attention}
A.~Vaswani, N.~Shazeer, N.~Parmar, J.~Uszkoreit, L.~Jones, A.~N. Gomez,
  {\L}.~Kaiser, and I.~Polosukhin, ``Attention is all you need,''
  \emph{Advances in neural information processing systems}, vol.~30, 2017.

\bibitem{li2017neural}
J.~Li, P.~Ren, Z.~Chen, Z.~Ren, T.~Lian, and J.~Ma, ``Neural attentive
  session-based recommendation,'' in \emph{Proceedings of the 2017 ACM on
  Conference on Information and Knowledge Management}, 2017, pp. 1419--1428.

\bibitem{kang2018self}
W.-C. Kang and J.~McAuley, ``Self-attentive sequential recommendation,'' in
  \emph{2018 IEEE international conference on data mining (ICDM)}.\hskip 1em
  plus 0.5em minus 0.4em\relax IEEE, 2018, pp. 197--206.

\bibitem{sun2019bert4rec}
F.~Sun, J.~Liu, J.~Wu, C.~Pei, X.~Lin, W.~Ou, and P.~Jiang, ``Bert4rec:
  Sequential recommendation with bidirectional encoder representations from
  transformer,'' in \emph{Proceedings of the 28th ACM international conference
  on information and knowledge management}, 2019, pp. 1441--1450.

\bibitem{wu2020sse}
L.~Wu, S.~Li, C.-J. Hsieh, and J.~Sharpnack, ``Sse-pt: Sequential
  recommendation via personalized transformer,'' in \emph{Fourteenth ACM
  Conference on Recommender Systems}, 2020, pp. 328--337.

\bibitem{fan2021continuous}
Z.~Fan, Z.~Liu, J.~Zhang, Y.~Xiong, L.~Zheng, and P.~S. Yu, ``Continuous-time
  sequential recommendation with temporal graph collaborative transformer,'' in
  \emph{Proceedings of the 30th ACM International Conference on Information \&
  Knowledge Management}, 2021, pp. 433--442.

\bibitem{fan2022sequential}
Z.~Fan, Z.~Liu, Y.~Wang, A.~Wang, Z.~Nazari, L.~Zheng, H.~Peng, and P.~S. Yu,
  ``Sequential recommendation via stochastic self-attention,'' in
  \emph{Proceedings of the ACM Web Conference 2022}, 2022, pp. 2036--2047.

\bibitem{kipf2017semi}
T.~N. Kipf and M.~Welling, ``Semi-supervised classification with graph
  convolutional networks,'' in \emph{Proceedings of International Conference on
  Learning Representations}, 2017.

\bibitem{luo2022clear}
X.~Luo, W.~Ju, M.~Qu, Y.~Gu, C.~Chen, M.~Deng, X.-S. Hua, and M.~Zhang,
  ``Clear: Cluster-enhanced contrast for self-supervised graph representation
  learning,'' \emph{IEEE Transactions on Neural Networks and Learning Systems},
  2022.

\bibitem{ju2023comprehensive}
W.~Ju, Z.~Fang, Y.~Gu, Z.~Liu, Q.~Long, Z.~Qiao, Y.~Qin, J.~Shen, F.~Sun,
  Z.~Xiao \emph{et~al.}, ``A comprehensive survey on deep graph representation
  learning,'' \emph{arXiv preprint arXiv:2304.05055}, 2023.

\bibitem{ju2023tgnn}
W.~Ju, X.~Luo, M.~Qu, Y.~Wang, C.~Chen, M.~Deng, X.-S. Hua, and M.~Zhang,
  ``Tgnn: A joint semi-supervised framework for graph-level classification,''
  \emph{arXiv preprint arXiv:2304.11688}, 2023.

\bibitem{wu2019session}
S.~Wu, Y.~Tang, Y.~Zhu, L.~Wang, X.~Xie, and T.~Tan, ``Session-based
  recommendation with graph neural networks,'' in \emph{Proceedings of the AAAI
  conference on artificial intelligence}, vol.~33, no.~01, 2019, pp. 346--353.

\bibitem{xu2019graph}
C.~Xu, P.~Zhao, Y.~Liu, V.~S. Sheng, J.~Xu, F.~Zhuang, J.~Fang, and X.~Zhou,
  ``Graph contextualized self-attention network for session-based
  recommendation.'' in \emph{IJCAI}, vol.~19, 2019, pp. 3940--3946.

\bibitem{yang2022multi}
Y.~Yang, C.~Huang, L.~Xia, Y.~Liang, Y.~Yu, and C.~Li, ``Multi-behavior
  hypergraph-enhanced transformer for sequential recommendation,'' in
  \emph{Proceedings of the 28th ACM SIGKDD Conference on Knowledge Discovery
  and Data Mining}, 2022, pp. 2263--2274.

\bibitem{qin2023disenpoi}
Y.~Qin, Y.~Wang, F.~Sun, W.~Ju, X.~Hou, Z.~Wang, J.~Cheng, J.~Lei, and
  M.~Zhang, ``Disenpoi: Disentangling sequential and geographical influence for
  point-of-interest recommendation,'' in \emph{Proceedings of the Sixteenth ACM
  International Conference on Web Search and Data Mining}, 2023, pp. 508--516.

\bibitem{luo2022dualgraph}
X.~Luo, W.~Ju, M.~Qu, C.~Chen, M.~Deng, X.-S. Hua, and M.~Zhang, ``Dualgraph:
  Improving semi-supervised graph classification via dual contrastive
  learning,'' in \emph{2022 IEEE 38th International Conference on Data
  Engineering (ICDE)}.\hskip 1em plus 0.5em minus 0.4em\relax IEEE, 2022, pp.
  699--712.

\bibitem{ju2023glcc}
W.~Ju, Y.~Gu, B.~Chen, G.~Sun, Y.~Qin, X.~Liu, X.~Luo, and M.~Zhang, ``Glcc: A
  general framework for graph-level clustering,'' in \emph{Proceedings of the
  AAAI Conference on Artificial Intelligence}, vol.~37, no.~4, 2023, pp.
  4391--4399.

\bibitem{ju2023unsupervised}
W.~Ju, Y.~Gu, X.~Luo, Y.~Wang, H.~Yuan, H.~Zhong, and M.~Zhang, ``Unsupervised
  graph-level representation learning with hierarchical contrasts,''
  \emph{Neural Networks}, vol. 158, pp. 359--368, 2023.

\bibitem{gilmer2017neural}
J.~Gilmer, S.~S. Schoenholz, P.~F. Riley, O.~Vinyals, and G.~E. Dahl, ``Neural
  message passing for quantum chemistry,'' in \emph{Proceedings of
  International Conference on Machine Learning}, 2017, pp. 1263--1272.

\bibitem{song2019session}
W.~Song, Z.~Xiao, Y.~Wang, L.~Charlin, M.~Zhang, and J.~Tang, ``Session-based
  social recommendation via dynamic graph attention networks,'' in
  \emph{Proceedings of the Twelfth ACM International Conference on Web Search
  and Data Mining}, 2019, pp. 555--563.

\bibitem{wang2022disenctr}
Y.~Wang, Y.~Qin, F.~Sun, B.~Zhang, X.~Hou, K.~Hu, J.~Cheng, J.~Lei, and
  M.~Zhang, ``Disenctr: Dynamic graph-based disentangled representation for
  click-through rate prediction,'' in \emph{Proceedings of the 45th
  International ACM SIGIR Conference on Research and Development in Information
  Retrieval}, 2022, pp. 2314--2318.

\bibitem{ju2022kernel}
W.~Ju, Y.~Qin, Z.~Qiao, X.~Luo, Y.~Wang, Y.~Fu, and M.~Zhang, ``Kernel-based
  substructure exploration for next poi recommendation,'' in \emph{2022 IEEE
  International Conference on Data Mining (ICDM)}.\hskip 1em plus 0.5em minus
  0.4em\relax IEEE, 2022, pp. 221--230.

\bibitem{qin2023diffusion}
Y.~Qin, H.~Wu, W.~Ju, X.~Luo, and M.~Zhang, ``A diffusion model for poi
  recommendation,'' \emph{arXiv preprint arXiv:2304.07041}, 2023.

\bibitem{velivckovic2018graph}
P.~Veli{\v{c}}kovi{\'c}, G.~Cucurull, A.~Casanova, A.~Romero, P.~Lio, and
  Y.~Bengio, ``Graph attention networks,'' in \emph{Proceedings of
  International Conference on Learning Representations}, 2017.

\bibitem{chen2018neural}
R.~T. Chen, Y.~Rubanova, J.~Bettencourt, and D.~K. Duvenaud, ``Neural ordinary
  differential equations,'' \emph{Advances in neural information processing
  systems}, vol.~31, 2018.

\bibitem{he2016deep}
K.~He, X.~Zhang, S.~Ren, and J.~Sun, ``Deep residual learning for image
  recognition,'' in \emph{Proceedings of the IEEE conference on computer vision
  and pattern recognition}, 2016, pp. 770--778.

\bibitem{fang2021spatial}
Z.~Fang, Q.~Long, G.~Song, and K.~Xie, ``Spatial-temporal graph ode networks
  for traffic flow forecasting,'' in \emph{Proceedings of the 27th ACM SIGKDD
  Conference on Knowledge Discovery \& Data Mining}, 2021, pp. 364--373.

\bibitem{choi2022graph}
J.~Choi, H.~Choi, J.~Hwang, and N.~Park, ``Graph neural controlled differential
  equations for traffic forecasting,'' in \emph{Proceedings of the AAAI
  Conference on Artificial Intelligence}, vol.~36, no.~6, 2022, pp. 6367--6374.

\bibitem{ji2022stden}
J.~Ji, J.~Wang, Z.~Jiang, J.~Jiang, and H.~Zhang, ``Stden: Towards
  physics-guided neural networks for traffic flow prediction,'' 2022.

\bibitem{rao2022fogs}
X.~Rao, H.~Wang, L.~Zhang, J.~Li, S.~Shang, and P.~Han, ``Fogs: First-order
  gradient supervision with learning-based graph for traffic flow
  forecasting,'' in \emph{Proceedings of International Joint Conference on
  Artificial Intelligence, IJCAI}.\hskip 1em plus 0.5em minus 0.4em\relax
  ijcai. org, 2022.

\bibitem{chen2011time}
Y.~Chen, B.~Yang, Q.~Meng, Y.~Zhao, and A.~Abraham, ``Time-series forecasting
  using a system of ordinary differential equations,'' \emph{Information
  Sciences}, vol. 181, no.~1, pp. 106--114, 2011.

\bibitem{de2019gru}
E.~De~Brouwer, J.~Simm, A.~Arany, and Y.~Moreau, ``Gru-ode-bayes: Continuous
  modeling of sporadically-observed time series,'' \emph{Advances in neural
  information processing systems}, vol.~32, 2019.

\bibitem{jin2022multivariate}
M.~Jin, Y.~Zheng, Y.-F. Li, S.~Chen, B.~Yang, and S.~Pan, ``Multivariate time
  series forecasting with dynamic graph neural odes,'' \emph{arXiv preprint
  arXiv:2202.08408}, 2022.

\bibitem{huang2020learning}
Z.~Huang, Y.~Sun, and W.~Wang, ``Learning continuous system dynamics from
  irregularly-sampled partial observations,'' \emph{Advances in Neural
  Information Processing Systems}, vol.~33, pp. 16\,177--16\,187, 2020.

\bibitem{huang2021coupled}
------, ``Coupled graph ode for learning interacting system dynamics.'' in
  \emph{KDD}, 2021, pp. 705--715.

\bibitem{luo2023hope}
X.~Luo, J.~Yuan, Z.~Huang, H.~Jiang, Y.~Qin, W.~Ju, M.~Zhang, and Y.~Sun,
  ``Hope: High-order graph ode for modeling interacting dynamics,'' 2023.

\bibitem{zhuang2019ordinary}
J.~Zhuang, N.~Dvornek, X.~Li, and J.~S. Duncan, ``Ordinary differential
  equations on graph networks,'' 2019.

\bibitem{xhonneux2020continuous}
L.-P. Xhonneux, M.~Qu, and J.~Tang, ``Continuous graph neural networks,'' in
  \emph{International Conference on Machine Learning}.\hskip 1em plus 0.5em
  minus 0.4em\relax PMLR, 2020, pp. 10\,432--10\,441.

\bibitem{wang2019neural}
X.~Wang, X.~He, M.~Wang, F.~Feng, and T.-S. Chua, ``Neural graph collaborative
  filtering,'' in \emph{Proceedings of the 42nd international ACM SIGIR
  conference on Research and development in Information Retrieval}, 2019, pp.
  165--174.

\bibitem{runge1895numerische}
C.~Runge, ``{\"U}ber die numerische aufl{\"o}sung von
  differentialgleichungen,'' \emph{Mathematische Annalen}, vol.~46, no.~2, pp.
  167--178, 1895.

\bibitem{xu2019self}
D.~Xu, C.~Ruan, E.~Korpeoglu, S.~Kumar, and K.~Achan, ``Self-attention with
  functional time representation learning,'' \emph{Advances in neural
  information processing systems}, vol.~32, 2019.

\bibitem{loomis2013introduction}
L.~H. Loomis, \emph{Introduction to abstract harmonic analysis}.\hskip 1em plus
  0.5em minus 0.4em\relax Courier Corporation, 2013.

\bibitem{rendle2012bpr}
S.~Rendle, C.~Freudenthaler, Z.~Gantner, and L.~Schmidt-Thieme, ``Bpr: Bayesian
  personalized ranking from implicit feedback,'' \emph{arXiv preprint
  arXiv:1205.2618}, 2012.

\bibitem{dormand1980family}
J.~R. Dormand and P.~J. Prince, ``A family of embedded runge-kutta formulae,''
  \emph{Journal of computational and applied mathematics}, vol.~6, no.~1, pp.
  19--26, 1980.

\bibitem{he2020lightgcn}
X.~He, K.~Deng, X.~Wang, Y.~Li, Y.~Zhang, and M.~Wang, ``Lightgcn: Simplifying
  and powering graph convolution network for recommendation,'' in
  \emph{Proceedings of the 43rd International ACM SIGIR conference on research
  and development in Information Retrieval}, 2020, pp. 639--648.

\bibitem{guo2022evolutionary}
J.~Guo, P.~Zhang, C.~Li, X.~Xie, Y.~Zhang, and S.~Kim, ``Evolutionary
  preference learning via graph nested gru ode for session-based
  recommendation,'' in \emph{Proceedings of the 31st ACM International
  Conference on Information \& Knowledge Management}, 2022, pp. 624--634.

\bibitem{li2020time}
J.~Li, Y.~Wang, and J.~McAuley, ``Time interval aware self-attention for
  sequential recommendation,'' in \emph{Proceedings of the 13th international
  conference on web search and data mining}, 2020, pp. 322--330.

\bibitem{wang2020disentangled}
X.~Wang, H.~Jin, A.~Zhang, X.~He, T.~Xu, and T.-S. Chua, ``Disentangled graph
  collaborative filtering,'' in \emph{Proceedings of the 43rd international ACM
  SIGIR conference on research and development in information retrieval}, 2020,
  pp. 1001--1010.

\bibitem{bao2021time}
J.~Bao and Y.~Zhang, ``Time-aware recommender system via continuous-time
  modeling,'' in \emph{Proceedings of the 30th ACM International Conference on
  Information \& Knowledge Management}, 2021, pp. 2872--2876.

\bibitem{li2023multi}
X.~Li, A.~Sun, M.~Zhao, J.~Yu, K.~Zhu, D.~Jin, M.~Yu, and R.~Yu,
  ``Multi-intention oriented contrastive learning for sequential
  recommendation,'' in \emph{Proceedings of the Sixteenth ACM International
  Conference on Web Search and Data Mining}, 2023, pp. 411--419.

\bibitem{yang2023debiased}
Y.~Yang, C.~Huang, L.~Xia, C.~Huang, D.~Luo, and K.~Lin, ``Debiased contrastive
  learning for sequential recommendation,'' in \emph{Proceedings of the ACM Web
  Conference 2023}, 2023, pp. 1063--1073.

\bibitem{van2008visualizing}
L.~Van~der Maaten and G.~Hinton, ``Visualizing data using t-sne.''
  \emph{Journal of machine learning research}, vol.~9, no.~11, 2008.

\end{thebibliography}
